\documentclass[aps,pre,twocolumn,superscriptaddress,floatfix]{revtex4}
\usepackage{dcolumn}
\usepackage{amsmath}
\usepackage{amssymb}
\usepackage{amsthm}

\usepackage{graphicx}
\usepackage{times}
\usepackage{bm}
\usepackage[T1]{fontenc}
\usepackage{color}
\usepackage{url}
\usepackage[bf]{subfigure}
\usepackage{rotating}
\usepackage{scalefnt}
\usepackage{multirow}
\usepackage{lineno}
\usepackage{setspace}
\usepackage{algorithm2e}
\usepackage{float}
\usepackage{natbib}
\usepackage{hyperref}
\usepackage{scalefnt}

\newcommand{\vertii}[1]{{\left\vert\kern-0.25ex\left\vert #1 \right\vert\kern-0.25ex\right\vert}}

\begin{document}

\title{Topological limits to parallel processing capability of network architectures}
\author{Giovanni Petri\footnote{Equal contribution}}
\email[]{Correspondence: giovanni.petri@isi.it}
\affiliation{ISI Foundation, Via Chisola 5, 10126 Torino, Italy}
\affiliation{ISI Global Science Foundation, New York, NY 10018, USA}

\author{Sebastian Musslick$^*$}
\affiliation{Princeton Neuroscience Institute, Princeton University, Princeton, NJ 08540, USA.}

\author{Biswadip Dey}
\affiliation{Siemens Corporation, Princeton, NJ 08536, USA.}

\author{Kayhan {\"O}zcimder}
\affiliation{MathWorks, Natick, MA, 01760 USA}

\author{David Turner}
\affiliation{Princeton Neuroscience Institute, Princeton University, Princeton, NJ 08540, USA.}

\author{Nesreen K. Ahmed}
\affiliation{Intel Labs, Santa Clara, CA 95054, USA.}

\author{Theodeore L. Willke}
\affiliation{Intel Labs, Santa Clara, CA 95054, USA.}

\author{Jonathan D Cohen}
\affiliation{Princeton Neuroscience Institute, Princeton University, Princeton, NJ 08540, USA.}
\affiliation{Department of Psychology, Princeton University, Princeton, NJ 08540, USA.}

\begin{abstract}
The ability to learn new tasks and generalize performance to others is one of the most remarkable characteristics of the human brain and of recent AI systems. 
The ability to perform multiple tasks simultaneously is also a signature characteristic of large-scale parallel architectures, that is evident in the human brain, and has been exploited effectively in more traditional, massively parallel computational architectures.  
Here, we show that these two characteristics are in tension, reflecting a fundamental tradeoff between interactive parallelism that supports learning and generalization, and independent parallelism that supports processing efficiency through concurrent multitasking.  
We formally show that, while the maximum number of tasks that can be performed simultaneously grows linearly with network size, under realistic scenarios (e.g. in an unpredictable environment), the expected number that can be performed concurrently at most grows strongly sub-linearly with network size. Hence, even modest reliance on shared representation, which supports more rapid learning and generalization, strictly constrains the number of tasks that can be performed simultaneously. This has profound consequences for understanding the human brain’s mix of sequential and parallel processing capabilities, as well the development of artificial intelligence systems that can optimally manage the tradeoff between learning and processing efficiency.  
\end{abstract}

\maketitle
There is a fundamental tension between two kinds of use for parallel distributed computing in network architectures.
The first focuses on incorporating a variety of interacting constraints in the learning and processing of complex representations (`interactive parallelism').
This has been profitably exploited in theories of human cognitive function  \cite{mcclelland1986appeal,rogers2004semantic} and most recently in the design of ``deep learning" artificial systems \cite{bengio2013representation,caruana1997multitask,baxter1995learning}.
In contrast, a second kind of use focuses on the capacity of a network to carry out multiple processes independently ('independent parallelism').  This approach has been exploited by massively parallel systems used in most modern computing clusters, and optimized by message-passing systems, such as MPI \cite{gropp1996high}, that seek to identify and distribute independent components of computation.  What has been less well explored is the relationship between these two types of parallelism, and the consequences that this has for the design and functioning of adaptive systems. 

Recent work has suggested that there is a fundamental tradeoff between these two types of parallelism that may help explain fundamental features of human cognitive function\cite{Musslick_et_al_2017}. In particular, human behavior presents an interesting puzzle with regard to multitasking ability.
On the one hand, we can effortlessly perform many kinds of tasks at the same time, such as walking, talking, and responding to our surroundings, all of which presumably involve extensive simultaneous computations. 
On the other hand, we are radically constrained in our ability to perform other kinds of tasks concurrently, such as planning a grocery list while simultaneously carrying out multidigit mental arithmetic.  
In cognitive psychology, this is attributed to a fundamental distinction between automatic and control-dependent processing \cite{posnerr,shiffrin1977controlled}. 
The former is capable of effortless, simultaneous execution, while the latter is subject to seriality constraints on performance. 

Early theorists proposed two alternative accounts for this constraint in control-dependent processing. One suggests that this reflects reliance on a centralized, limited capacity mechanism (akin to the core of a traditional computer), thus explaining the dramatic limitation in the human ability to simultaneously perform multiple control-dependent tasks.  
The alternative interpretation suggests that constraints in control-dependent processing reflect the purpose, rather than a limitation, of control mechanisms: to resolve conflicts that arise from competition among the resources required to perform specific combinations of tasks, which themselves rely on the shared use of representations \cite{wickens1991processing,allport1980attention,meyer1997computational,navon1979economy}.  

While compelling, the latter proposal was not undergirded by formal analysis of the extent to which shared use of representations constrains processing at the system level. 
In particular, one concern might be that shared use of representations in a system as large as the human brain may pose minimal constraints on parallel processing. 
Recently, however, numerical work has shown that even modest sharing of representations among tasks can impose radical constraints on simultaneous execution due to cross-talk interference among tasks, and that the effects of such interference can be invariant to network size.  For example, Feng et al. \cite{Feng_et_al_2014} and Musslick et al. \cite{Musslick_et_al_2016} found that simple models with modest sharing of representations between processing pathways showed a severe reduction of processing capacity that was virtutally invariant over the number of pathways in the network. 
However, these models relied on numerical simulations, and were thus constrained in the size of networks that they could address. 

Understanding the source of such constraints, and explaining them explicitly in mechanistically and formally rigorous terms remains an important challenge not only for understanding human performance -- and how it arises from computations in the brain -- but also for the design of artificial systems that can emulate human performance.

Here we provide a general theoretical analysis of the problem, based in a combination of graph theory and statistical mechanics of frustrated systems. 
We illustrate the mechanism by which even modest degrees of shared representations impose strong constraints on the number of tasks that can be performed simultaneously without the risk of interference from crosstalk between tasks. 
Our results highlight a fundamental tension in network architectures between the benefits that accrue from shared representations (i.e. flexibility of processing and generalization \cite{bengio2013representation,caruana1997multitask,baxter1995learning}) and their cost in terms of processing efficiency (i.e., the number of independent tasks that can be performed in parallel \cite{Musslick_et_al_2017}).


	\section*{Results}
\subsection*{Measures of task dependency predict parallel processing capability in a trained neural network}
To consider the problem of multitasking (i.e., concurrent parallel processing) analytically, we first provide a formal definition of a task. 
For more details, see Section S1 in the Supplementary Information (SI). 
Given an input space $I$ of stimuli (e.g. colors) and an output space $O$ of responses (e.g. verbal response), a task $T: I \to O$ represents a mapping between the two (e.g., naming the color of a stimulus), such that the mapping is independent of any other, and that selection of a feature from its input space can be made independently of any other. 
Different tasks can share an input space, output space, or both (e.g., reading a color word such as "red" out loud, and naming the color in which it is printed, share an output space).  When this occurs, there is the potential for the tasks to interfere with one another \cite{stroop1935studies,Feng_et_al_2014}. 

\begin{figure*}[ht!]
\centering 
\includegraphics[width=\textwidth]{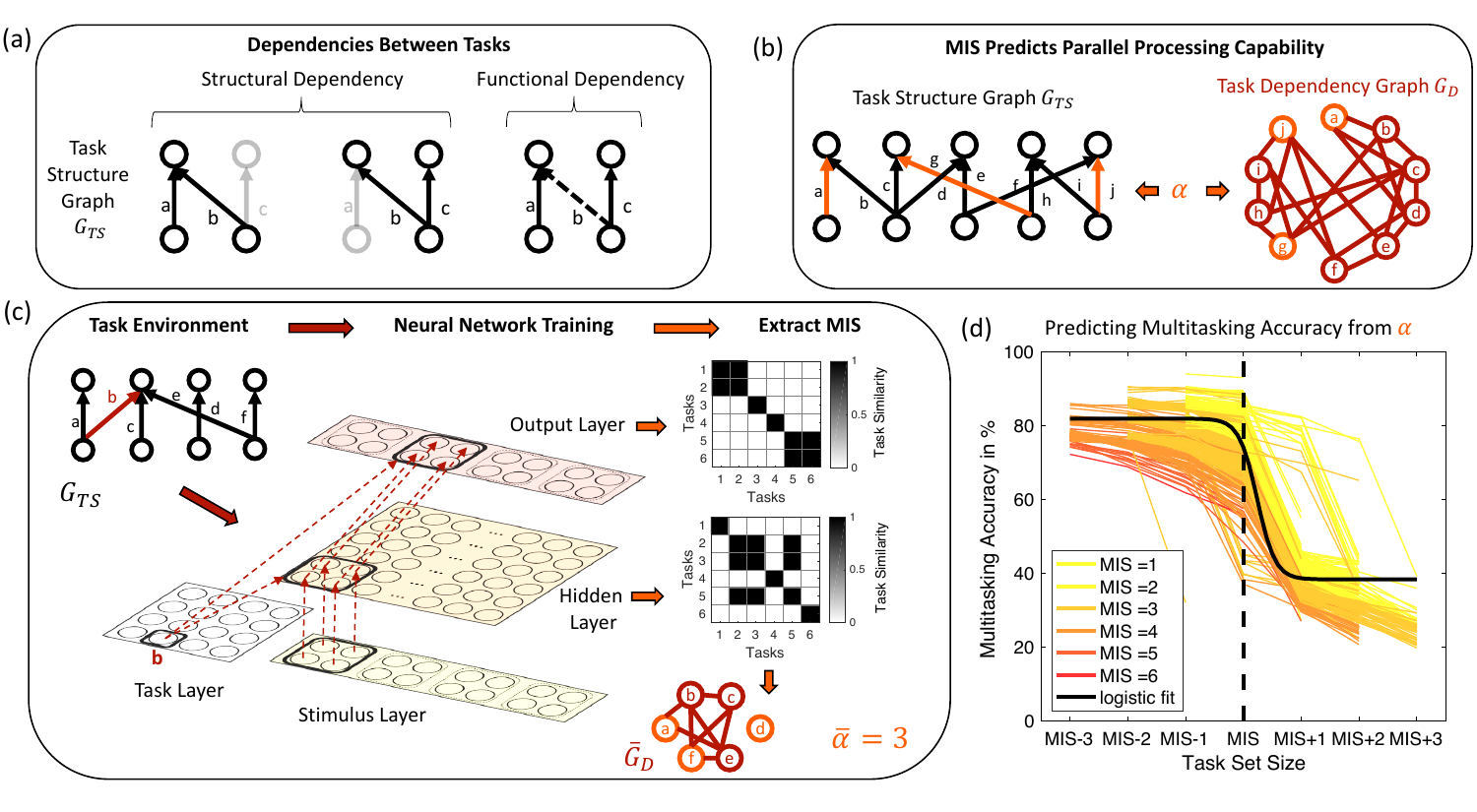}
\caption{\textbf{Graph-theoretic measures predict parallel processing capacity.} (a) A bipartite task structure graph $G_{TS}$ describes tasks in terms of mappings from an input space to an output space. Each task corresponds to an edge in $G_{TS}$ from an input node to an output node. 
Two tasks are: i) structurally dependent if they converge to the same output node (\textit{left}) or originate from the same input node (\textit{middle}); ii) functionally dependent if their edges are connected by a third task (\textit{right}). 
(b) Dependencies between tasks in $G_{TS}$ are expressed by the corresponding task dependency graph $G_D$: tasks are now represented as nodes, and they are connected if the two tasks are structurally or functionally dependent. The MIS of $G_D$ corresponds to the largest set of tasks that a network can execute in parallel without interference\cite{Musslick_et_al_2016}. Its cardinality $\alpha = |MIS|$ is the maximal parallel processing capacity of the network $G_{TS}$. (c) A neural network is trained on a set of tasks, in which each task requires the network to map a set of features from the stimulus layer via a hidden layer to a set of features on the output layer. Each task is designated by a unit in an additional (task) input layer that projects to both the hidden and output layers. All tasks in the environment can be expressed in terms of a task structure graph $G_{TS}$ (as shown in (a)).
The network is trained on all tasks by activating, on each trial, a particular task unit and an input unit corresponding to a stimulus feature in the set for that task, and requiring the network to activate the corresponding output unit.  The average activity patterns at the hidden and output layers across all inputs under a given task are taken as the network's representation of that task.  
The two resulting similarity matrices (for the hidden and output layers) are used to infer dependencies between tasks based on shared task representations, and to construct the empirical task dependency graph $\bar{G}_D$, which we use to predict the empirical $\bar{\alpha}$.  
(d) The parallel processing capacity of the trained network is predicted by $\bar{\alpha}$. The plot shows the highest multitasking accuracy of the network as a function of the number of tasks it is asked to perform in parallel (performance curve) as indicated in relation to the network's MIS. Each line corresponds to the multitasking performance of a trained network, whereas the color of each line indicates the predicted MIS for that network. The solid black line depicts the average fit of a logistic function to accuracy curves across networks. 
} 
\label{fig::panel1}
\end{figure*}

Such interference can be made explicit by describing the task structure in the form of a (bipartite) task structure graph $G_{TS} = G(\mathcal{I}, \mathcal{O}, \mathcal{T})$. 
$G_{TS}$ makes the sharing of representations across tasks explicit (Figure \ref{fig::panel1}a), in which $\mathcal{I}$, $\mathcal{O}$ and $\mathcal{T}$ are respectively the sets of input spaces, output spaces and tasks. 
A task $t \in \mathcal{T}$ is formally defined as mapping from an input space to an output space $t: I \to O$, with $I\in\mathcal{I}, O\in\mathcal{O}$.
Whenever two tasks share an input node $I$ or an output node $O$ we assume that they are at risk of interference due to direct cross-talk and therefore should not be executed in parallel; we call this dependency \textit{structural} because of the direct reliance on common resources \cite{Musslick_et_al_2016}. 
Figure \ref{fig::panel1}a depicts this type of dependency between tasks $a-b$ and between $b-c$. 
Importantly, in addition to structural dependence, there can also be \textit{functional} dependence between two tasks: this is the case whenever, given two tasks, a third task maps the input space (i.e. connects the input node) of the first task to the output space (i.e. output node) of the second one. 
In Figure ~\ref{fig::panel1}a, tasks $a$ and $c$ are functionally dependent via task $b$, because activating a stimulus in task $c$'s input space does the same for $b$, thus invoking a response to $b$ that may conflict with the response to $a$.

Finding the maximum number of tasks that can be simultaneously executed (i.e., multitasked) is then equivalent to finding the largest set of edges in $G_{TS}$ that are neither structurally nor functionally dependent on one another.  
In graph-theoretic terms, this corresponds to finding a maximum induced edge matching of $G_{TS}$: a subset of tasks in which none of the tasks either share a node or are connected by an edge.
In Figure~\ref{fig::panel1}b (left) we show an example of induced matching (in orange).

Interestingly, under an assumption that we will specify in detail shortly, all task dependencies can be made explicit in a derived graph, the \textit{task dependency graph} $G_D$, in which nodes represent tasks, and edges their (structural or functional) dependencies (Figure \ref{fig::panel1}b, right).
Starting from $G_{TS}$, the dependency graph is built by considering the square of the line graph of $G_{TS}$. 
In fact, the line graph of $G_{TS}$ encodes structural interferences between tasks. 
Taking its square correponds to closing all open wedges and encodes functional interferences. 
It can be shown that  the maximum induced edge matching on $G_{TS}$ corresponds to the maximum independent set (MIS) of $G_D$ \cite{Gavril_1973}, the largest set of nodes that are not connected by any edge.
The cardinality of this set is called the independence number $\alpha$ of $G_D$. 

This equivalence is key to our first main contribution: a neural network constrained to learn a task structure characterized by graph $G_{TS}$ exhibits a maximum parallel capacity given by the independence number of the corresponding $G_D$. 

To assess the correspondence of this theoretical measure of parallel processing capacity to the performance of an actual network, we trained a simple non-linear feed-forward network (see Figure \ref{fig::panel1}c), with four layers, that has been used previously to simulate a wide array of empirical findings concerning human cognitive performance \cite{cohen1990control, cohen1992parallel, botvinick2001conflict}. 
The network architecture entails two input layers, one that encodes the current stimulus (stimulus layer) and another one that encodes the task to be performed on the stimulus (task layer). 
Both input layers project to a hidden layer that computes an internal representation of task-relevant input features of the stimulus. 
Finally, information encoded at the hidden layer is projected together with the task layer input to an output layer at which the response of the network is computed. The weight projections from the task layer serve to bias processing towards task-relevant stimulus information represented at the hidden layer, as well as task-relevant responses at the output layer \cite{cohen1990control}. This, in turn, shapes the representations of the input and output space respectively for each task.  

As a benchmark, we first demonstrate the correspondence between the structure of $G_{TS}$ and the one derived from the theoretical dependence graph $G_D$ under assumption of maximum sharing of representations.  To achieve this, we train a set of networks to learn the mapping from inputs to outputs for each task, with fixed weight projections from the task layer to the hidden layer that are the same for tasks with shared input spaces.  This guarantees the maximum amount of representation sharing between tasks. We refer to this as a minimal basis set representation, as it is the most compact form of representation at the hidden layer that can support performance of all tasks.  We trained 400 networks in this manner, varying the total number of tasks (between $4-30$) and task structure graph $G_{TS}$ (Figure \ref{fig::panel1}c).
For each network trained on a task environment $G_{TS}$, we computed a theoretical task dependency graph $G_D$ (see SI). 


Figure \ref{fig::panel1}d  shows that $\alpha$ predicts the maximum number of tasks the network can perform in parallel.
That is, the highest accuracy that the network can achieve (across all task combinations for a given task set size) drops as soon as task set size exceeds $\alpha$. 
To statistically evaluate this prediction we fit a logistic function to the accuracy of a network's performance as a function of set size. We find that the inflection point of the sigmoid curve is accurately predicted by the $\alpha$ derived from $G_D$. That is, the inflection point (i.e., offset) of the curve lies significantly above a set size equal to $\alpha$ , $t(352) = 9.1465, p < 10^{-17}$, and below a set size of $\alpha+1$, $t(352) = -24.3986, p < 10^{-77}$. These predictions turn out to be robust for a range of different performance metrics, number of hidden units in the network, as well as choices of $\theta$ that is used to extract $\bar{G}_{D}$ (see Methods and Section S.3).

These results show that, when the network is constrained to learn maximally shared representations, the pattern of performance it exhibits is consistent with the task interference structure described by $G_{TS}$, and that its parallel processing capacity can be accurately predicted from the corresponding $G_D$, which is easily obtained from $G_{TS}$.

These results validate the graph analysis with network simulations under conditions of maximal sharing — that is, when the network is constrained to use the minimal basis set representation. However, they do not address other network configurations that do not conform to the minimal basis set.  These are of interest for theoretical and practical reasons.  
While the minimal basis set maximizes representational efficiency and generalization \cite{Musslick_et_al_2017,saxe2019mathematical, caruana1997multitask}, it constrains multitasking capability by introducing functional dependence among tasks.  
This can be mitigated by the use of separated representations and, in the limit, a dedicated set of representations for each task (i.e., combination of input space and output space) —- a scheme we refer to as the tensor production representation.

Previous work has used empirical simulations to examine how multitasking performance fares between the extremes of minimal basis set and tensor product representations \cite{sagiv2018efficiency,MusslickCohen2019}. 
However this has been constrained to relatively small networks, because doing so in larger networks becomes computationally intractable: the number of multitasking conditions that must be evaluated grows factorially with the number of tasks the network can perform.  This comports with the graph representation of the problem, where it is known that computing the maximum independent set for the square of a line graph is NP-hard \cite{berman1994approximating,tarjan1977finding}. The methods proposed here offer an analytic approach to this problem, but this first requires that they be validated for network configurations that extend beyond the minimal basis set.  Such validation would not only license their use in addressing theoretical issues, but also be of practical value by allowing multitasking performance to be estimated from network (and corresponding neural) measurements that can be performed on individual tasks rather than all combinations of them, thus scaling linearly rather than factorially with the number of tasks. 

A possible solution to this problem would be to directly investigate the effects of network structure on $\alpha$ computing it from the neural network itself. 
However, this is computationally prohibitive, because for a neural network with $N$ nodes it requires the enumeration of all node subsets and thus scales as $O(N!)$.
Using the MIS size computation affords a significant computational advantage, as the algorithmic complexity of computing the independence number explicitly for a $G_D$ with $M$ nodes is $O(1.2^{M})$ and with $M$ scales as $O(N^2)$ in our case \cite{bollobas1998modern,Tarjan_SJC_77}. 
However, although efficient algorithms for specific classes of graphs exist \cite{Quaddoura14,Perfect_MIS_Poly_1984,Gavril_SJC_72}, measuring $\alpha$ directly from $G_D$ remains impractical for large graphs relevant to most real-world applications. 
More importantly, as mentioned above, algorithmic solutions do not provide insights about the features of the task structure that are responsible for limiting parallel processing capacity. 
To obtain these, we need an analytical solution based on the properties of $G_{TS}$ and $G_D$.

In the following two sections, we address these issues. 
First, we relax constraints on weights from the task layer to the hidden layer, allowing the network to learn hidden layer representations that range between the minimal basis set and tensor product configurations, making it difficult if not impossible to determine $G_{TS}$ from the weights. We show that multitasking performance can instead be inferred from task-related patterns of activity over the hidden and output layers of the network. We also demonstrate that even when this is the case, multitasking performance can be derived from $G_{TS}$. Second, we show that, even when $G_{TS}$ is not accessible, the parallel processing capacity of the network can be accurately predicted from an analytic expression based solely on the degree distributions of an empirical $\bar{G}_D$ inferred from patterns of activity.

\subsection*{Relaxing the basis set assumption.}
\begin{figure*}
\centering
\includegraphics[width=\textwidth]{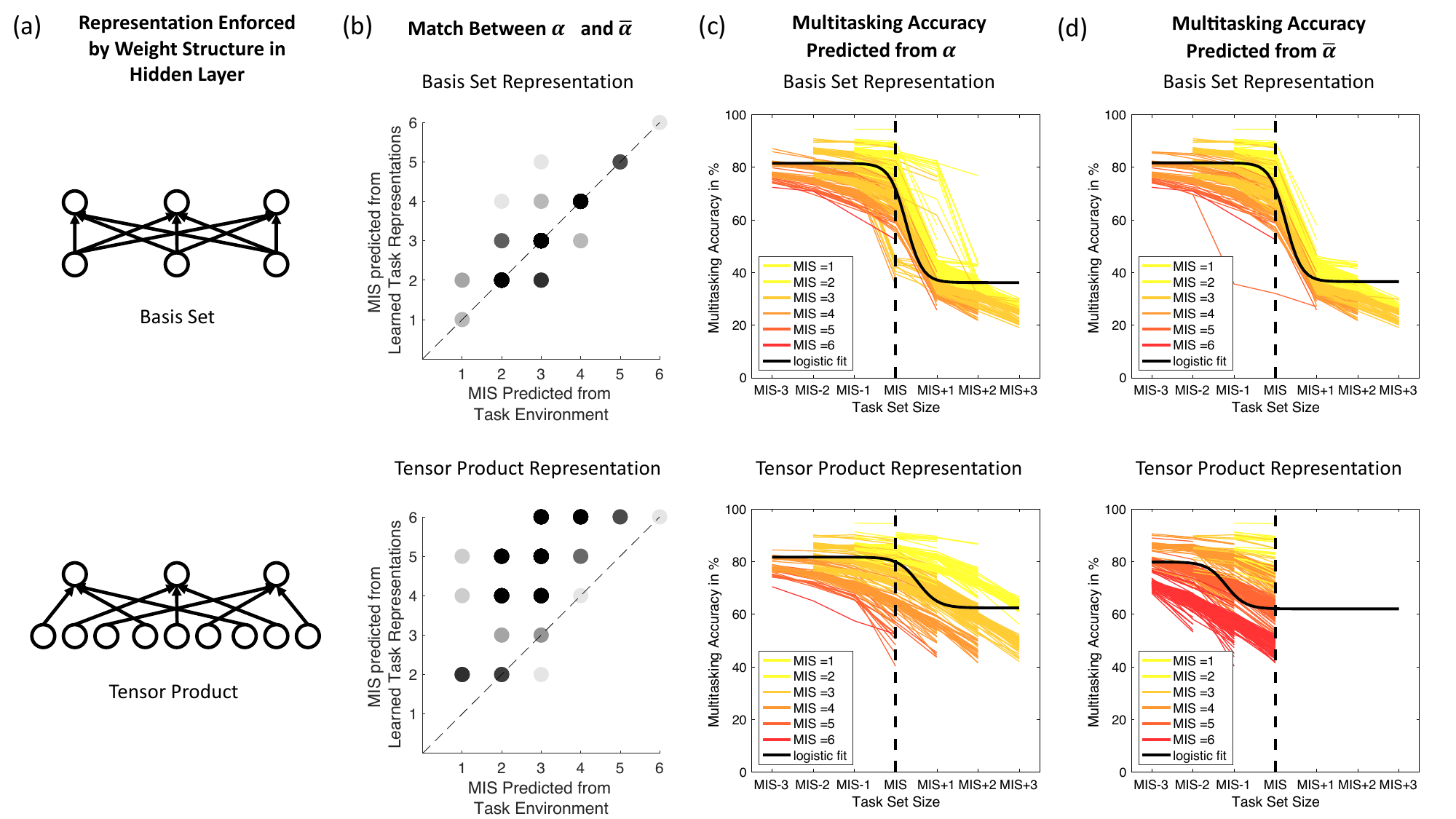}
\caption{\textbf{Impact of the weights learning scheme on $\alpha$ and $\bar{\alpha}$}. (a) An exemplary task environment with nine tasks (three input dimensions and three output dimensions) may be learned in two different ways, depending on the weight structure of the network. If the weight structure enforces a basis set representation, all tasks sharing the same input dimension share a representation in the hidden layer of the network (first layer of the extracted bipartite graph). If a tensor product representation is enforced, each of the nine tasks is dedicated a separate representation in the hidden layer of the network. (b) Match between the MIS predicted from the task structure graph, $\alpha$, and the MIS predicted from learned representations extracted from the neural network, $\bar{\alpha}$. The darker the color of a dot, the higher the proportion of networks representing this configuration. (c) Parallel processing capacity of the trained network predicted by $\alpha$. (d) Parallel processing capacity of the trained network is predicted by $\bar{\alpha}$. Plots in (c, d) show the highest multitasking accuracy of the network as a function of the number of tasks it is asked to perform in parallel (performance curve) as indicated in relation to the network’s MIS. Each line corresponds to the multitasking performance of a trained network, whereas the color of each line indicates the predicted MIS for that network. The solid black line depicts the average fit of a logistic function to accuracy curves across networks. 
}
\label{fig:weights-scheme}
\end{figure*}

Here we compare the minimal basis set configuration (Fig \ref{fig:weights-scheme}a top) and tensor product configuration (Fig \ref{fig:weights-scheme}a bottom). 
With the latter, we observe that the network does not learn the same task structure as the one provided by $G_{TS}$, and hence the corresponding learned dependency graph $\bar{G}_D$ does not correspond to the theoretical one computed starting from $G_{TS}$ (Figure \ref{fig:weights-scheme}b). 
In particular, to show this, we trained 400 neural networks\footnote{We increased the number of hidden units to 500 to accommodate the tensor product scheme.} under the minimal basis set and the tensor product schemes. 
In every instance, we specified a task structure graph $G_{TS}$ and computed its associated theoretical $G_D$. 
For comparison, we also constructed an empirical $\bar{G}_D$ by examining the similarity between single task representations encoded in the hidden and output layers \cite{Musslick_et_al_2016}. 
For each single task, we computed the average activity vector generated at the hidden and output layers by all inputs for that task, and used that as the representation of that task at each of those layers.  Two tasks were considered to be \emph{structurally} dependent if their representations at either the hidden or output layer exceeded a Pearson correlation threshold of $\theta = 0.5$, indicating that they shared either common associative and/or output representations respectively.  A pair of tasks was considered to be \emph{functionally} dependent if there was a third task for which the hidden layer representation was similar to one task in the pair and the output layer representation was similar to the other task in the pair. We then computed the theoretical capacity $\alpha$ from $G_D$ and the empirical capacity $\bar{\alpha}$ directly from $\bar{G}_D$ (see Methods, Section S2 and S4).

Consistent with the results reported above, for the minimal basis set configuration we find nearly perfect agreement between $\alpha$, $\bar{\alpha}$ and the network capacity measured directly by the inflection point in the network multitasking performance (Figure \ref{fig:weights-scheme}b-c-d top). The inflection point of multitasking accuracy predicted by $\bar{\alpha}$ was significantly above $\bar{\alpha}$, $t(354) = 8.9606, p < 10^{-17}$, and below $\bar{\alpha}+1$, $t(354) = -26.5022, p < 10^{-85}$. 
For the tensor product configuration, the predictions of $G_D$ and $\bar{G}_D$ diverge (Figure \ref{fig:weights-scheme}b bottom). 
In particular, we find that the $\alpha$ obtained from the theoretical $G_D$ predicts maximum parallel processing capacity well. The inflection point of the fitted logistic curve lies significantly above $\alpha$, $t(367) = 13.2679, p < 10^{-32}$, and below $\alpha+1$, $-11.0636, p < 10^{-24}$. In contrast, estimating the maximum capacity $\bar{\alpha}$ empirically (i.e., from network activations) provides a more liberal prediction of the network's multitasking performance (Figure \ref{fig:weights-scheme}c bottom). While the inflection point of the fitted sigmoid lied significantly below $\bar{\alpha}+1$, it does not significantly lie above $\bar{\alpha}$, indicating that the multitasking performance of the network is slightly over-predicted by the maximum capacity extracted from neural network activations. Predicted theoretical parallel processing capacity $\alpha$ was found to be lower on average ($M= 2.8338, SD = 0.8424$) compared to the empirical multitasing capacity $\bar{\alpha}$ ($4.7375, SD = 1.1143$). 
We also find that predictions obtained from  $\bar{G}_D$ are less noisy, yet on average nearly equivalent to predictions obtained from the theoretical $G_D$  (Section S4). 
Critically, this shows that estimates of $\bar{G}_D$ derived strictly from single task measurements yield a good estimate of the network's MIS, without requiring direct access to the underlying $G_{TS}$.

This observation has important practical significance. Whereas technologies such as fMRI and EEG are available to measure task-related patterns of neural activity, there is still no way to reliably estimate task-specific patterns of connectivity that would be required to directly estimate $G_{TS}$. Furthermore, since direct assessment of multitasking capability grows factorially with the number of tasks under consideration, doing so behaviorally is impractical in all but the most limited task environments.  However, since the number of measurements required to estimate $\bar{G}_D$ from patterns of activity grows only linearly with the number of tasks, it should be possible to use this (i.e., using neural measures of task-specific patterns of activity) to estimate parallel processing capacity in settings involving realistic numbers of tasks that would otherwise be impossible.

Finally, in the SI (Section S5) we show that the same results hold for more complex neural networks, for example ones with multiple hidden layers. In this case, we can compute a $\bar{G}_D$ for each pair of consecutive layers and use the smallest $\bar{\alpha}$ across all layers as the bound on the network parallel processing capacity. This assumption is justified by the fact that effectively the pair of layers with the largest interference acts as a bottleneck for the whole network. We find good agreement between this prediction and the measured network accuracy (Figure S.5).

\begin{figure*}
\centering 
\includegraphics[width=\textwidth]{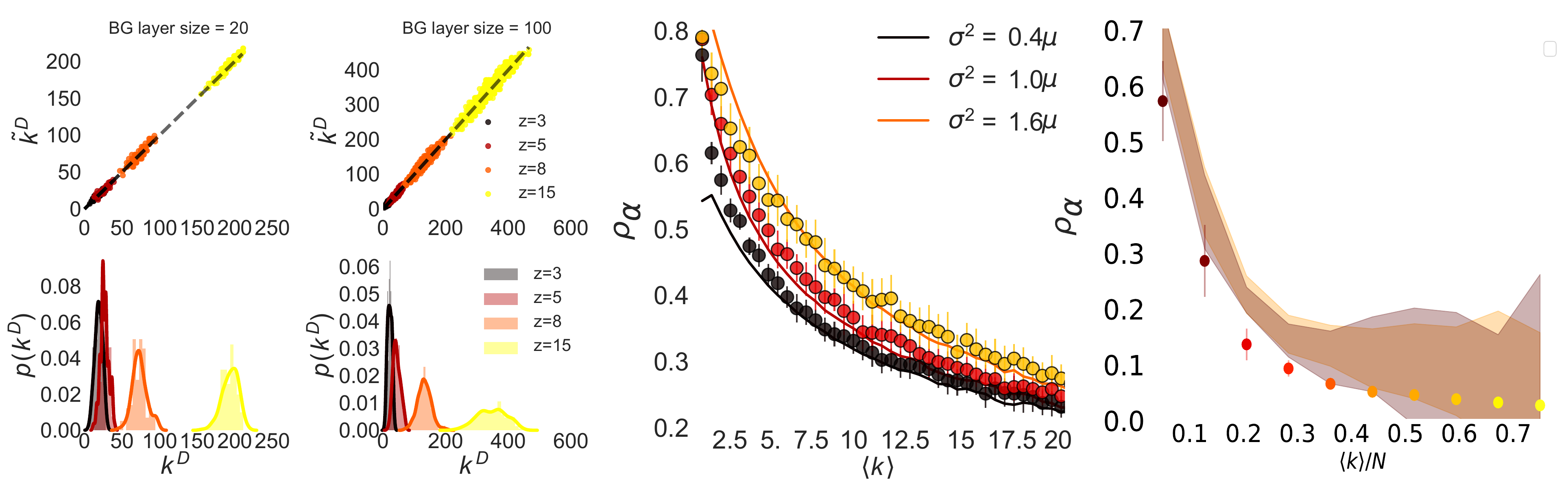}
\caption{\textbf{Graph-theoretic results for $\rho_\alpha$ and $\rho_\phi$.}
\textbf{(a)} Comparison of estimated $\bar{k}^D$  and actual $k^D$ for dependency graphs obtained from task structure graphs with a range of average degrees $z$. For each value $z$, the task dependency graph $G_D$ was computed from its bipartite $G_{TS}$ with a fixed degree on the input nodes, and a binomial distribution of degrees for the output nodes (similar to an Erdös-Rényi graph, and following \cite{Feng_et_al_2014}).
The lower plot shows the same information in distribution form, with discrete bins corresponding to $\bar{k}^D$ and the solid line corresponding to the actual distribution. 
\textbf{(b)} MIS densities $\rho_\alpha$ for a set of generic networks with Gaussian degree distributions of varying widths, comparing the exact computation (dots) with the values predicted from Equation \ref{eq:rhoalpha} (solid lines) as a function of the Gaussian distribution mean $\mu$. The plots show that, for fixed $\mu$, increasing the degree heterogeneity ($\sigma^2$) is associated with increased $\rho_\alpha$.
\textbf{(c)} MIS densities $\rho_\alpha$ for $G_D$ as a function of the task structure graph density $\langle k \rangle N$, comparing i) the explicit calculation on the theoretical $G_D$ (dots); ii) the analytical results using Equation \ref{eq:rhoalpha} and the $G_D$ degree distribution estimated using Equation \ref{eq:dependency-degree} (pink shading); and iii) the analytical results using Equation \ref{eq:rhoalpha} with the measured $G_D$ degree distribution. We created a set of $G_{TS}$ graphs with binomial degree distributions on the input and output layers. For each parameter choice, we computed 50 $G_{TS}$ graphs. Error bars on dots correspond to one standard deviation of the resulting $\rho_\alpha$ values. The color of the dots represents the average density of the corresponding $G_D$d, which ranged from 0.2 to 0.7. The shaded areas are one standard deviation intervals for the predicted $\rho_\alpha$s.}\label{fig::results}
\end{figure*}

\begin{figure*}
\centering 
\includegraphics[width=\textwidth]{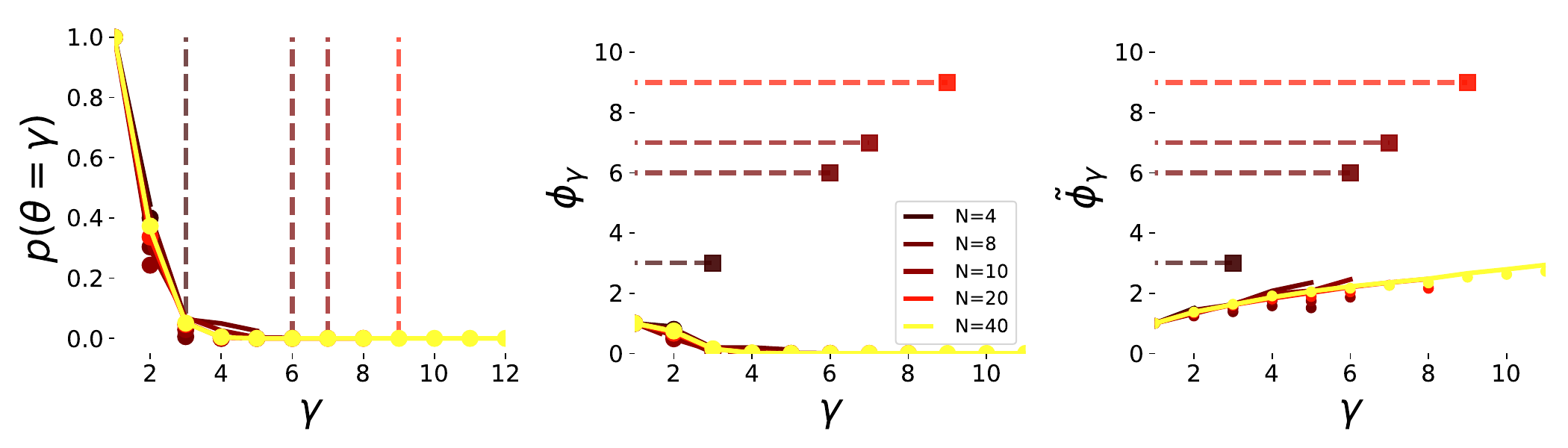}
\caption{\textbf{Graph-theoretic results for $p_\gamma$, $\phi_\gamma$ and $\tilde{\phi}_\gamma$.}
\textbf{a)} Probability $p_\gamma$ that all tasks in a task subset of cardinality $\gamma$ are performed successfully (i.e. are independent). We plot $p_\gamma$ measured directly (dots) in dependency graphs obtained from Erdos-Renyi task graphs with density $\rho=0.2$ and variable size ($N=4,8,10,20,40$) and compare them with the predictions of Eq.~\ref{prob:beta:gamma} (solid lines). The theoretical $p_\gamma$ are slightly higher than the measured values, but overall we find good agreement between the two. Vertical dashed lines highlight the MIS values for various $N$ and show how for all sizes the probability of randomly choosing a maximal independent task set is vanishingly small. 
\textbf{(b)} The $\phi_\gamma$ values (obtained using the $p_\gamma$ in panel $(a)$) are much lower than the MIS for the corresponding size (shown as the horizontal dashed lines); moreover, $\phi_\gamma$ displays little dependence on the network size $N$ as opposed to the MIS, which grows linearly instead. 
\textbf{(c)} Even under the more permissive reward function,  $\tilde{\phi}_\gamma$ only grows sublinearly with $\gamma$ and again shows no dependence on the $G_{TS}$ size $N$, leading to strongly diminishing returns for the effective parallel capacity. 
}\label{fig::success-results}
\end{figure*}

\subsection*{Maximum parallel capacity estimation for dependency graphs of arbitrary size} 
The results above are promising, but they were validated in small networks.  An important theoretical question that remains unanswered is how the relationship between sharing of representations and parallel processing capacity of a network scale with network size for very large networks, for which it is prohibitive to compute either $\alpha$ and $\bar{\alpha}$ directly.

To address this problem, we develop a graph ensemble formulation of the MIS problem in terms of the degree distribution of the task dependency graph. 
This allows us to tease apart the roles of graph density and heterogeneity, independently of the network size, and to make general observations about the relationship between task dependency and task encoding policies in determining parallel processing capacity.  
To achieve this, we need to relate $\rho_\alpha$ ($\rho_{\bar{\alpha}}$) to the degree distribution of $G_D$ ($\bar{G}_D$).

When considering the minimal basis set configuration, the degree distribution of $G_D$ can be computed directly starting from $G_{TS}$. This can be done in a manner similar to the standard calculation of the number of second neighbors \cite{Newman:2010:NI:1809753}. 
Since the task dependency graph $G_D$ is the square of the line graph $\mathcal{L}(G_{TS})$ of $G_{TS}$,  the estimated degree $\bar{k}_e^D$ of task $e$ in $G_D$ as $\bar{k}_e^D$:
\begin{eqnarray}\label{eq:dependency-degree}
\bar{k}^{D}_e  &  \simeq    &\frac{(k_i-1)\langle k_o^2 \rangle}{\langle k_o\rangle} +  \frac{(k_o-1)\langle k_i^2 \rangle}{\langle k_i\rangle} \\  
& & - \frac{(k_i-1)(k_o-1)(\langle k_i \rangle -1) (\langle k_o \rangle -1)}{M-1} \nonumber
 \end{eqnarray}
where $\langle \bullet \rangle$ are the expectation values of $k_i$ and $k_o$, and $M$ is the number of edges in $G_{TS}$ (or equivalently of nodes in $G_D$). We refer the reader to Section  S.9 for full details.

In Figure \ref{fig::results}a (and the SI) we show that Eq.~\ref{eq:dependency-degree} gives good results (Pearson's $R>0.9$) for graphs of various densities and for various degree distributions. 
Note that $\bar{k}^{D}$ is written in terms of the first two moments of $p_{st}$, recovering the previously observed connection between the heterogeneity of $G_{TS}$ graph and that of the corresponding dependency graph $G_D$. 

When not in the minimal basis set scheme, it is not possible in general to obtain an expression for the degree distribution of $\bar{G}_D$ from $G_{TS}$. However, this is a minor limitation since, as explained in the previous section, it is possible to estimate $\bar{G}_D$ from the network activations, even when $G_{TS}$ is unknown. Thus, going forward, we will focus exclusively on dependency graphs, disregarding their origin (theoretical or empirical). For simplicity of notation, we will denote these as $G_D$ and their independence number as $\alpha$, dropping the $\bar{\bullet}$.

Equipped with the dependency graph's degree distribution, we now need to estimate the expected maximum independence number $\alpha$.
We do this by building on recent work by Lucibello and Ricci-Tersenghi \cite{Lucibello:2014ga} and estimate the independence number density $\rho_\alpha = \alpha/M$ (where $M$ is the number of nodes in $G_D$), based on a factor graph description of the maximum set packing problem, of which the independence number problem is a particular instance.
Crucially, these expressions allow $\rho_\alpha$ to be related directly to the graph's structure, and in particular the graph's degree distribution, thus providing a window into the role of the network's topology. 
 
Exploiting the properties of the degree and excess degree (the degree a node reached following an edge) distributions \cite{newman2009random}, the $\rho_\alpha$ estimate can be rewritten in terms of degree generating functions, and takes the form: 
\begin{equation}\label{eq:rhoalpha}
\rho_\alpha = \frac{\langle k \rangle}{\langle c \rangle} \left( 1 - p_*^{c/(c-1)} \right) + [M_k(t) - M_k'(t)]
\end{equation}
\noindent where $t=\log p_*$, and $p_*$ needs to satisfy the self-consistent equation 
\begin{equation}\label{eq:pstar}
p_*  = \mathbb{E}_{\tilde{c}}\left( 1 - \frac{1}{\langle k \rangle p_*} M_k'(t) \right)^{\tilde{c}}. 
\end{equation}
Here $k$ is the node degree in $G_D$, $c$ and $\tilde{c}$ refer to the factor nodes' degrees and excess degrees, which in the case of the MIS are fixed to $c=2$, $\tilde{c}=1$ (see Methods and SI), and $M_k(t)$ is the generating function for the degree distribution $p(k)$. 
\emph{In the case of the minimal basis set configuration}, Eq. \ref{eq:dependency-degree} can be used to calculate the degree distribution of $G_D$ from $G_{TS}$, which, substituted in Eq. \ref{eq:rhoalpha}, yields the estimate for $\rho_\alpha$. 
\emph{For the unconstrained case}, one can directly use the degrees from the empirical $\bar{G}_D$.  \\ 

For certain classes of graphs, for which we have analytical degree generating functions, it is possible to obtain insights into the role of the density and heterogeneity of $G_D$ directly. 
For example, for a Gaussian distribution with mean $<k>$ and variance $\sigma^2$,  the moment generating function takes the form $M_k(t) = e^{\langle k \rangle t + \sigma^2 t^2/2}$. 
Substituting the expression above, we obtain
\begin{equation}
p_* = \left [1 - \frac{1}{\langle k \rangle p_*} (\langle k \rangle + \sigma^2 \ln p_*) M_k(\ln p_*) \right]
\end{equation}
Solving numerically for $p_*$, we can then compute:
\begin{equation}
\rho_\alpha = \frac{\langle k \rangle}{\langle c \rangle} \left( 1 - p_*^{c/(c-1)} \right) + M_k(\ln p_*) (1 - \langle k \rangle - \sigma^2 \ln p_*)
\end{equation}

In Figure \ref{fig::results}b  we show that this expression provides a close approximation of the behaviour of $\rho_\alpha$ for increasing network density and for various levels of degree heterogeneity of $G_D$. 
Importantly, it provides an analytical grounding for the previous empirical observations that increased heterogeneity of task overlap for a given average density results in a higher $\rho_\alpha$ \cite{Feng_et_al_2014,Musslick_et_al_2016}. Here, we used Gaussian degree distributions to explicitly illustrate the impact of the density of the dependency graph, which depends, in turn, on  the density of the task structure graph and its degree heterogeneity: 
for a fixed size, dense and uniform graphs have a smaller MIS than sparse, heterogeneous ones.

Finally, imagining we are in the minimal basis set scheme, we show that it is possible to predict $\rho_\alpha$ starting from the degree distribution of $G_{TS}$, estimating the degree distribution of $G_D$ from it, and then plugging it in Eq. \ref{eq:rhoalpha}. 
We computed this for a set $G_{TS}$ with fixed number of nodes per layer $N$ and for increasing densities (see SI for other $G_{TS}$ topologies).
We find that the prediction obtained from the esitmated and actual degree distributions of $G_D$ are in agreement, and they both yield a strict upper bound on $\rho_\alpha$. 

\subsection*{Effective parallel processing capacity}
Above, we showed that information contained in the degree distribution of $G_D$ can be used to provide a good approximation of its independence density $\rho_\alpha$, and hence the corresponding independence number $\alpha_G = \rho_\alpha  M$ for networks of arbitrary size.

However, the independence number is specific to a particular subset (or very few subsets) of tasks.
That is, that maximum level of parallelism can only be achieved by performing the particular combination of tasks belonging to a maximum independent set. While this describes an upper bound on the parallel processing capacity of the network, it does not take account of the likelihood, given the environment, that all of the particular tasks in a maximum independent set are available to be performed at a given time (i.e., all of the relevant stimuli are present and corresponding motor affordances viable).  Thus, the independent numbrer does not address the more practical question, that is more likely to be relevant in naturalistic settings: what is the greatest number of tasks that the system be expected to perform simultaneously \textit{on average}, given a probability distribution of tasks in the environment.  In other words, given a task set $T$ of cardinality $|T| = \gamma$, what is the probability $p_\gamma$ that those tasks are  both available for execution and can be successfully executed at the same time? 

The probability $p_\gamma$ is a special case of the probability $p(\theta, \gamma, G_D)$ of successfully executing $\theta$ out of $\gamma$ tasks from a dependency graph $G_D$. 
The latter requires the $\theta$ nodes in $T \subset G_D$ not to be linked with each other, and the remaining $\gamma - \theta$ nodes to be connected to at least one of the first $\theta$ tasks. 
For a graph $G_D$, we can estimate the probability of successfully executing  $1\leq\theta\leq\gamma$ tasks in $T$ as:
\begin{equation} \label{prob:beta:gamma}
p(\theta;\gamma, \mathcal{G}_D) \simeq  \left(1 - \frac{\langle k^2 \rangle}{2M_D} \right)^{\binom{\theta}{2}} \left(\frac{\theta \langle k \rangle^2}{2M_D}\right)^{\gamma - \theta}
\end{equation} 
where $M_D$, $\langle k \rangle$ and $\langle k^2 \rangle$ are respectively the number of edges in $G_D$, the first and the second moment of $G_D$'s degree distribution. 

In Figure \ref{fig::success-results}a we show the value of $p_\gamma$ as a function of $\gamma$, for various $G_D$ with variable network size $M$ but fixed network density. 
Naturally for $\gamma=1$, the probability of executing the task is always 1 since a single task cannot interfere with itself, but $p_\gamma$ decreases very rapidly as the number of attempted tasks increases and, remarkably, does not depend on the network size (see SI for details). 
Equation \ref{fig::success-results} confirms analytically the size independence of the MIS previously observed in numerical experiments by Feng et al. \cite{Feng_et_al_2014}: at fixed density for $G_D$, $\langle k \rangle$, $\langle k^2 \rangle$ and $M_D$ all scale as $M^2$, making Equation \ref{fig::success-results} independent of $M$. 

To quantify how the precipitous decrease in $p_\gamma$ relates to the outcome of performance, we associate a reward with each multitasking attempt. 
We consider two reward schemes: in the first (\textit{all-or-nothing}) scheme, we give a positive reward to an attempt to perform $\gamma$ tasks only if all tasks are successful (i.e. independent in $G_D$) and no reward otherwise; in the second (\textit{graded}) scheme, we give a reward to each multitasking attempt on $\gamma$ tasks proportional to the maximum number of independent tasks $\gamma'\leq\gamma$ in that set.  

These schemes encode two extremes in how rewards for performance might depend on multitask success. 
The all-or-nothing scheme corresponds to situations in which the outcomes of the tasks can influence one another, and thus all tasks need to be successfully performed (e.g. juggling a collection of objects requires all individual objects to be successfully juggled; failing on one is likely to induce failure on the others).
In contrast, the graded scheme corresponds to situations in which task outcomes are not correlated (e.g. driving and listening to a conversation) and hence failing one task does not induce failure of others.
For the former, the expected reward is therefore written, modulo a multiplicative coefficient, as 
\begin{equation}\label{eq:phi}
\phi(\gamma, G_D) = \gamma p_\gamma + 0(1-p_\gamma) = \gamma p_\gamma, 
\end{equation}  
which peaks at low values of $\gamma$ and rapidly converges to zero (Figure \ref{fig::success-results}b).

For the more permissive graded scheme, we have:
\begin{equation}\label{eq:tilde-phi}
\tilde{\phi}(\gamma, G_D) = \sum_{\theta=1}^{\gamma} p(\theta, \gamma, G_D) \theta.
\end{equation}
In this case, $\tilde{\phi}$ grows for increasing $\gamma$ values (Fig \ref{fig::success-results}c). 
This is expected because, under this scheme, for any task subset the reward is positive (e.g. at the limit if the task set is the whole network, the reward is proportional to the MIS size). 
Despite this, the average reward $\tilde{\phi}$ grows sublinearly with $\gamma$ and, again, does not depend on the network size. 

As a consequence of this sublinear increase, any increase in dependency graph size is associated with diminishing returns in both $\phi$ and $\tilde{\phi}$. 
In the SI (Figures S.11-12), we show that, in the case of fixed average degree of $G_D$, the effective parallel processing capacity weakly depends on $M$, but the qualitative results do not change. 
Finally, we show that also in the small $M$ limit (outside the regime of validity of the formal treatment), a qualitatively similar effect can be observed for the empirical effective parallel processing capacity of trained neural networks (Figure S.13).

\section*{Discussion}

The work presented provides a formal analysis of the idea
that the two forms of parallelism described here are not merely differences in computational strategy, but reflect a fundamental computational trade-off in network architectures between efficiency of learning and generalization vs. efficiency of processing: the very network fabric that supports interactive parallelism by sharing representations between tasks (e.g. for learning and/or generalization) induces limits on independent parallelism and processing efficiency-- that is, their ability to perform multiple tasks simultaneously \cite{Musslick_et_al_2017}.

This trade-off may seem obvious.  However, a quantitative characterization of its consequences, an understanding of how a system as large and complex as the human brain manages it, and the implications it has for the design of artificial systems have not yet been explored in any depth.  
Previous work using numerical simulations has suggested that the impact of shared representations is severe and asymptotically invariant with the size of the network \cite{Feng_et_al_2014,Musslick_et_al_2016}.  
However, the problem can be shown to be NP-hard, and thus computational tractability limited those studies to relatively small networks.  
Here, we presented an approximation of the problem that is analytically tractable, and thus can be used to examine it at arbitrary scales, permitting an analysis of its manifestation in more complex systems - both natural and artificial.

More specifically, we modeled the structure of cross-talk between tasks as graphs and studied their properties borrowing techniques from network science and from statistical mechanics of frustrated systems. 
This allowed us to to summarize how the topology of the task dependency graph (i.e. of its degree distribution) affects the neural architecture's parallel capacity.  Furthermore, we validated the use of this method when the underlying graph structure of the network is estimated from individual patterns of task-specific activity, that can be measured empirically (e.g., from neural data) and that scales linearly with the number of tasks rather than factorially with the number of task combination (as would be required for direct assessment of multitasking capability from behavioral performance; we return to this issue below).

We provided analytic confirmation that the reduction is extreme when there is only modest sharing of representations across tasks, and that this applies at all scales. 
The latter is something that could not be confirmed in numerical simulations.  
We also found that inducing heterogeneity in the connectivity of the task structure graph (for fixed density) or in the task dependency graph itself relieves some of the constraints induced by shared representations, resulting in greater parallel processing capacity.

However, again, our analytic treatment allowed us to freely examine the effect of scale, where we observed that the benefit scales in a strikingly sublinear manner, with rapidly decreasing returns in parallel capacity with network size, even when the proportion of shared representations (and attendant rate of competition) is kept constant. 

The network models we analyzed here used direct mappings between inputs and outputs.  
We showed that the phenomenology in a simple example of deeper network is the same, which suggests that our results will be relevant to more realistic, multilayer networks, such as those used in deep learning applications.
In fact, the definition of a task used in our analyses applies in such networks as well, as each pair of layers can be considered as an input-output mapping, and thus the entire network can be considered as a series of such single-layered networks. 
From this perspective, a task to be executed by the network as a whole (i.e., a mapping from its initial input to its final output) can be decomposed into a series of subtasks, traversing the various layers. While this allows for the task to be successfully reproduced by multiple paths, at the same time the likelihood of interference between pathways implementing different tasks increases with the number of intermediate layers (i.e., opportunities for intersection), compounding the effects we have described for single layer networks \cite{alon2017graph,alon2018multitasking}. 
The same logic applies to recurrent networks.  These factors are similar to the effects of path structure on controllability in unfolded temporal graphs \cite{posfai2014structural,li2017fundamental}. 

Our work also provides the basis for developing methods of assessing the parallel processing capacity of natural agents (e.g,. humans) for a given set of tasks.  
Previously, methods have been proposed for doing so that use explicit signal modeling to infer the parallel processing capacity of a system from behavioral data (reaction time distributions) generated by actual task performance \cite{townsend2004theory,townsend2004serial,wenger2006costs}.  
However, these measures are constrained in two respects.  
First, they are indirect, inferring parallel vs. serial processing from patterns of performance.  Second, they require that performance be measured for all combinations of tasks over which parallel processing capacity is to be estimated.  Thus, the number of measurements required grows factorially with the number of tasks to be considered.  The methods we report here complement this approach, and may help address its limitations.  Like behavioral measures, they provide a means for estimating parallel processing capability when the underlying task structure is not known, but in this case using measurement of the representations engaged by each individual task, such as  the patterns of neural activity measured using direct neuronal recordings or neuroimaging methods such as fMRI or EEG.  The similarities among the patterns of activity elicited by each individual task (e.g., correlations among them) can be used to construct a task dependency graph - analogous to how we constructed dependency graphs from correlations in the patterns of activity in the synthetic neural networks - which, in turn, can be used to estimate the parallel processing capacity of the agent.  Since this requires only that the pattern of activity be measured for each individual task, and not all combinations of tasks, the measure scales linearly rather than factorially with the number of tasks to be evaluated.  

This may be valuable for important real world domains, where multitasking is critical, but in which it is impractical to individually and exhaustively evaluate all of the potential combinations of tasks that may need to be performed - for example pilots who must monitor a large number of instruments, each of which requires a response on a different device, and all of which must be executed as quickly as possible -- ideally simultaneously.  At present, there are no systematic methods for reliably and exhaustively evaluating which or how many combinations of tasks a pilot is able to perform at once - this is done empirically.  Our methods, combined with neuroimaging methods, may provide a practical means for carrying out such evaluations and thereby improving both skill assessment and training.

One potential limitation of the proposed analysis is that averages of task representations need not necessarily reflect the extent to which sharing of representations occurs across tasks, which must occur at the level of individual stimulus features. While the results presented in this, as well as other work \cite{Musslick_et_al_2016,MusslickCohen2019} suggest that task averages do seem to provide a good proxy for the similarity in representations across tasks, others have reported work on using geometric measures to identify the manifolds on which representations for different tasks live in neural networks \cite{bernardi2018geometry,cohen2019separability}. It remains a matter for future research to explore how well these measures can be used to predict predict parallel processing capacity of network architectures.

At a higher level of analysis, our methods may also help shed light on how a system balances the efficiency of learning and generalization provided by interactive parallelism and shared representations, at the cost of serial processing, with the efficiency of processing provided by independent parallelism, at the cost of greater training time and task-specificity.  
This is a question that has been at the heart of research on skill acquisition, and the relationship between control-dependent vs. automatic processing in psychology and neuroscience for over half a century. To date, these are questions that have been addressed largely in qualitative terms, or using descriptive models.  Here, we provide a first formal analysis of this tradeoff, in terms of the constraints that interactions among processes (shared representations) impose on the capacity for independent parallelism, and  that allows the tradeoff between the two forms of processing to be explicitly quantified.  
This tradeoff also poses a control and meta-learning problem, that is likely to be of interest not only to psychologists and neuroscientists interested in how people manage this trade-off, but also to machine learning efforts to design artificial systems that match or exceed the balance of flexibility and efficiency of human capabilities.  
At present, the two approaches to parallel computation are generally treated as independent design alternatives, with interactive parallelism (e.g., deep learning networks) used to address problems that require the simultaneous satisfaction of many constraints (such as face recognition or natural language processing; \cite{lecun2015deep,caruana1997multitask,bengio2013representation,baxter1995learning}), and independent parallelism (e.g., traditional multicore and large-scale compute clusters) are used for problems that can be broken up into isolated, independent processes.  
In contrast, the human brain appears to have integrated the use of both forms of parallelism - a feature that is likely to be fundamental to its capacity for adaptation - with interactive parallelism affording it the ability for generalization and inference, and independent parallelism for multitasking in domains requiring processing efficiency.  
For example, people continually make decisions about whether to engage in performing a task as quickly but not as proficiently as possible (e.g., hunt-and-peck at a computer or musical keyboard), or to expend the time and effort to acquire a more fluid form of skill that relies on parallel execution of actions (e.g., touch typing, or playing chords on a musical instrument).  Recent work has posed this as a meta-learning problem, and begun to explore algorithms that estimate the future discounted value of learning a task quickly using shared representations, but at the expense of control-dependence and a seriality constraint, versus expending the additional time to learn task-dedicated representations that afford multitasking capability and thus greater processing efficiency  \cite{Musslick_et_al_2017,sagivefficiency,Ravi_inPrep}. The analyses we report here may help lend formal rigor to future extensions of such work.  This may, in turn, have  relevance for system design in a wide range of other domains, such as maximum channel capacity for coding problems  \cite{butenko2002finding}, jobs with interfering schedules \cite{bomze1999maximum,wan2011wireless} and register allocation \cite{hack2006register}. 

At present, the approach we described is limited in a number of ways, however it provides a clear direction for addressing these. For example, we used undirected binary dependency graphs, corresponding to the assumption that interference between processes is symmetric and all or none. In natural systems, of course, interactions can be both asymmetric and graded, a feature that can be captured by the use of directed weighted graphs. The generalization of our methods to such graphs presents considerable challenges, and is an important direction for future research.  Nevertheless, the correspondence of our theoretical results with the numerical analyses (implementing asymmetric and graded forms of interference) suggests that our findings may provide useful approximations for current applications, and a valuable foundation for future theoretical work.


\section*{Methods}
\textbf{Neural network architecture and processing.}
We use a standard non-linear feed-forward network, with four layers, that has been used previously to simulate a wide array of empirical findings concerning human cognitive performance \cite{cohen1990control, botvinick2001conflict, rogers2004semantic}. The network consists of two input layers, one of which represents the stimulus presented to the network and another that encodes an instruction for the task that the network has to perform on this stimulus. Both input layers project to a hidden layer. Unless stated otherwise, the hidden layer contained 100 units.  Both the hidden layer and the task layer further project to an output layer that computes the network's response. The real-valued activity of each input unit constitutes the current stimulus. Activated units in the task layer indicate the task(s) to be currently executed. Performing a single task corresponds to clamping the corresponding task unit to 1 (activated) while all other units are set to 0. Multitasking conditions are represented by activating multiple task units at the same time. Units in the hidden and output layers take values between 0 and 1, as determined by a logistic activation function applied to their net input.
Stimulus input units are structured according to $D$ dimensions (subvectors of the stimulus pattern), each of which is comprised of a set of $D$ feature units with only one feature unit activated per dimension. Similarly, output units are organized into $D$ response dimensions, with only one of the $D$ response units permitted to be active within a response dimension. Each task is represented by a single task input unit that is associated with a set of unique, one-to-one mappings between the input units in one stimulus dimension and the output units in one response dimension, and that is independent of the mappings for all other tasks (see Figure \ref{fig::panel1}c). The number of stimulus input dimensions and response dimensions was varied between 4 and 9 across environments.  The task mappings were generated with the Erd\H{o}s-R\'{e}nyi model such that the number of overlapping tasks for a given stimulus input dimension $z$ varied between 1 and 7. For each environment $G_{TS}$ a network was initialized with a set of small random weights and then trained using the backpropagation algorithm \cite{rumelhart1986david} to produce the task-specified response for all stimuli in each task until it reached a mean-squared error performance of 0.001.\footnote{The training criterion was chosen such that the network achieves single task performance comparable to that of human participants on tasks requiring simple stimulus-response mappings (accuracy \%).} We constrained the learned representations of the network to reflect the task similarity structure of the environment $G_{TS}$ by fixing the weights from the task units to the hidden layer: Weight vectors for tasks relying on the same stimulus input dimensions were set to yield a Pearson correlation coefficient of value $1$ whereas weight vectors for tasks of non-overlapping stimulus dimensions were uncorrelated.

\textbf{Dependency graph extraction.}
We follow the analysis described in \cite{Musslick_et_al_2016} and focus on the representations (patterns of activity) over the hidden and output units, insofar as these reflect the computations carried out by the network required to perform each task. In particular, we are interested in the characteristics  of these representations for each task, how they compare across tasks, and how these factors correspond to empirically evaluated parallel processing performance. The representations associated with each task can be characterized by calculating, for each unit in the hidden and output layers, the mean of its activity over all of the stimuli for a given task;  this mean pattern of activity can then be used as a representation of the task.

Correlating patterns of activity within a layer across tasks yields a task similarity matrix that can be examined separately for the hidden and output layers of the network.  This can then be used to assess the extent to which different tasks rely on similar or different representations within each layer of the network.  Figure~\ref{fig::panel1}c provides an example of such similarity matrices (thresholded for similarity correlations above $\theta = 0.5$).  Tasks that have similar representations over the hidden layer can be inferred to rely on the same input dimension —-- that is, they share an input component in the bipartite graph representation of the network —-- and tasks that are similar at the output layer can be inferred to share an output component.  Accordingly, a task dependency graph $\bar{G}_D$ (of the type shown in Figure~\ref{fig::panel1}b) can be constructed by measuring the patterns of activity observed in the network while it performs each individual task. \\

\textbf{Assessing Multitasking Accuracy.} To test the overall multitasking performance for each network, we considered all sets of “multitaskable" tasks on which it was trained; that is, all sets of structurally independent tasks, for which each task had input and output dimensions that were distinct from all of the others in the set.  The accuracy of the network on a single task was determined by the probability of responding correctly in the task-relevant output dimension, averaged across all stimuli. Multitasking accuracy for a given set of tasks was determined by the average probability of responding correctly across all task-relevant output dimensions, averaged across all stimuli. The probability of responding correctly in a given output dimension was determined by a leaky competitive accumulator layer \cite{usher2001time} (optimized for performance under each multitasking condition), implementing the assumption that the network could only provide one response per response dimension (see SI for details). To statistically assess predictability of multitasking performance, we fit a logistic curve to the best multitasking accuracy as a function of set task set size. To avoid ill-conditioned solutions for logistic fits, we excluded networks for which the number of data points fell below 3. Multitasking accuracy is considered well predicted if the inflection point (bias) of the fitted logistic lies significantly above the predicted task set size and significantly below the predicted set size + 1. 

\subsection*{Code and Data Availability Statement}
Code and data are available at \url{https://github.com/lordgrilo/Multitasking_capacity}.

\section*{Acknowledgements}
G.P. received funding support from Fondazione Compagnia San Paolo and from Intesa Sanpaolo Innovation Center. 

\section*{Author contributions}
GP and SM contributed equally to this paper. GP, SM, BD, KO, NKA, TW and JDC designed research; GP developed and performed analytical and numerical calculations; SM and DT designed, implemented and performed  the neural network simulations; SM, KO, BD, NKA provided tools and performed neural network analysis ; JDC and TW conceptualized research, provided advice for all parts of the work; GP, SM, BD, KO, NKA, TW and JDC wrote the paper. 

\section*{Conflicts of interest}
The authors declare no conflicts of interest exist.


\widetext
\pagebreak
\begin{center}
\textbf{\large Supplemental Materials: Topological limits to parallel processing capability of network architectures}
\end{center}
\setcounter{equation}{0}
\setcounter{figure}{0}
\setcounter{table}{0}
\setcounter{page}{1}
\makeatletter
\renewcommand{\theequation}{S\arabic{equation}}
\renewcommand{\thesection}{S\arabic{section}}   
\renewcommand{\thetable}{S\arabic{table}}   
\renewcommand{\thefigure}{S.\arabic{figure}}
\renewcommand{\bibnumfmt}[1]{[S#1]}
\renewcommand{\citenumfont}[1]{S#1}

\section{Definition of a Task and Multitask}

In the main text, we define a task as input space $I$ of stimuli (e.g. colors) and an output space $O$ of responses (e.g. verbal response). Here, we elaborate on what we mean by those terms and illustrate how they translate into valid definitions of a multitask. 

We define an input space as any set of inputs values, each of which corresponds to a different value of an output space for a given task, and therefore only one of which can legally be present at a given time for the execution of that task. Similarly, an output space is a set of output values each of which corresponds to a different input value within the input space of a given task, and only one of which can be executed at a given time.  Note that these definitions are relative to the definition of a task. These are intentionally abstract definitions, that serve as the basis for the formal treatment here and in related work.  

Although these definitions are abstract, they capture what is usually meant by a task in the context of behavioral studies, where it is typically defined by the mapping from a set of mutually exclusive stimuli to a set of mutually exclusive responses. For instance, consider a classic task environment, the Stroop paradigm \cite{stroop1935studies}, in which stimuli consist of color words displayed in a particular color (e.g. the word "GREEN" displayed in red). Participants may either be asked to name the color in which the stimulus is displayed (color naming) or to read the word out loud (word reading). The input space of the color naming task constitutes the set of all possible displayed colors whereas the input space of the word reading task constitutes the set of all possible color words. The input spaces of the two tasks in this environment entail different modalities (colors and words). In a different task environment,  input spaces may correspond to different feature dimensions within a modality. For example, an experimenter may ask participants to either name the hue of a  color, or name its saturation, thus relying on input spaces involving different feature domains (saturation and hue) within same modality (stimulus color). In this example, the task environment involved only one output dimension (verbal responding) but it could be extended to include others (such as manual responding). For example, participants could be asked to read the word and press a corresponding button. This illustrates that the definition of input spaces and output spaces is always relative to the definition of the tasks (i.e. mappings between input spaces and output spaces) specified by the environment.
 
In accordance with definitions in cognitive psychology \cite{kahneman1973attention, pashler1994dual, Feng_et_al_2014}, we assume that two or more tasks constitute a valid multitask if none of the tasks shares an input space or output space with another task. Thus, the definition of input spaces and output spaces determine what can be treated as a multitask. To illustrate this, consider an illegal case in which two tasks sharing the same input space map to two different output spaces with responses that are not incompatible with one another -- for example, reading a word out loud, and pointing in a word-specific direction (e.g., point left for "RED", right for "GREEN", etc.).  It is of course easy to imagine learning to do this rather easily.  However, this does not fall within a definition of a genuine multitask, insofar as the latter requires that the inputs for each of the tasks involved in a multitask be drawn \emph{independently} of one another.  This can't be accomplished using the same input space for two different tasks, this would frequently violate the assumption that two different values can be represented within the same input space at the same time -- except in cases in which the different tasks all \emph{happen to draw the same value} in their shared input space.  Another way of thinking about this is that if tasks that share the same input space always also use the same input \emph{value} from that space (e.g., all Stroop stimuli were \emph{always} congruent), then they are not really \emph{independent} tasks; rather, they can be thought of as \emph{one} task with a more complex output space (i.e., with compound responses).  Thus, for these reasons, we exclude tasks that share the same input space from participating in legal multitasks.

\section{Assessing task performance using Leaky Competitive Accumulator}
We assessed performance of each task with an LCA layer\cite{usher2001time} that was assigned to each response dimension. The LCA layer was comprised of a set of units $r_i$ that received as their input the activity of corresponding units in that response dimension. The winning response was determined by the accumulation of activity by each LCA unit, and the competition among them, the dynamics of which were given by
\begin{equation}
dr_i = [y_o - \lambda r_i + \alpha_{\text{LCA}} f(r_i)  - \beta \sum_{j \neq i} f(r_j)] \frac{dt}{\tau} + \xi_i \sqrt{\frac{dt}{\tau}}
\end{equation}

where $y_o$ is the activity of the corresponding response unit in a given response dimension, $\lambda$ is the decay rate of $r_i$, $\alpha_{\text{LCA}}$ is the recurrent excitation weight\footnote{Note that $\alpha_{\text{LCA}}$ is different from $\alpha$ in the main text. While the former has been used to denote recurrent excitation in the LCA units\cite{usher2001time}, the latter has been used notate the maximum parallel processing capacity of a network\cite{alon2017graph}.} of $r_i$, $\beta$ is the inhibition weight between LCA units, $\tau$ is the rate constant, and $\xi$ is noise sampled from a Gaussian distribution with zero mean and standard deviation $\sigma$.  The activity of each LCA response unit was lower bounded by zero via a threshold such that $f(r_i) = r_i$ for $r_i \geq 0$ and $f(r_i) = 0$ for $r < 0$. The response for a given response dimension was determined by the unit within the corresponding LCA layer the activity $f(r_i)$ of which first reached threshold $z$. The accuracy for each response dimension  corresponded to the probability of generating the correct response for that dimension $P(correct)$ across 100 simulations of the LCA, and the reaction time $RT$ for that dimension was the average number of time steps  required for the response to reach threshold. The following parameter values were used for all reported simulations: $\lambda = 0.4$, $\alpha_{\text{LCA}} = 0.2$, $\beta = 0.2$, $\sigma = 0.1$, and $z$ for each LCA layer was chosen as the threshold that maximizes reward rate ($P(correct)/RT$) for that dimension.

\clearpage
\section{Robustness of Network Analysis}

Our neural network analysis involved
extracting an empirical dependency graph $\bar{G}_D$ from single task representations. Two tasks were considered structurally dependent if their average activity vectors in the hidden or output layer of the network were correlated above some threshold $\theta$. We then computed the empirical MIS $\bar{\alpha}$ of $\bar{G}_D$ and tested whether the best multitasking accuracy of the network drops as soon as the number of tasks that the network was asked to multitask exceeded $\bar{\alpha}$. 

\begin{figure}
\centering
\includegraphics[width=1\textwidth]{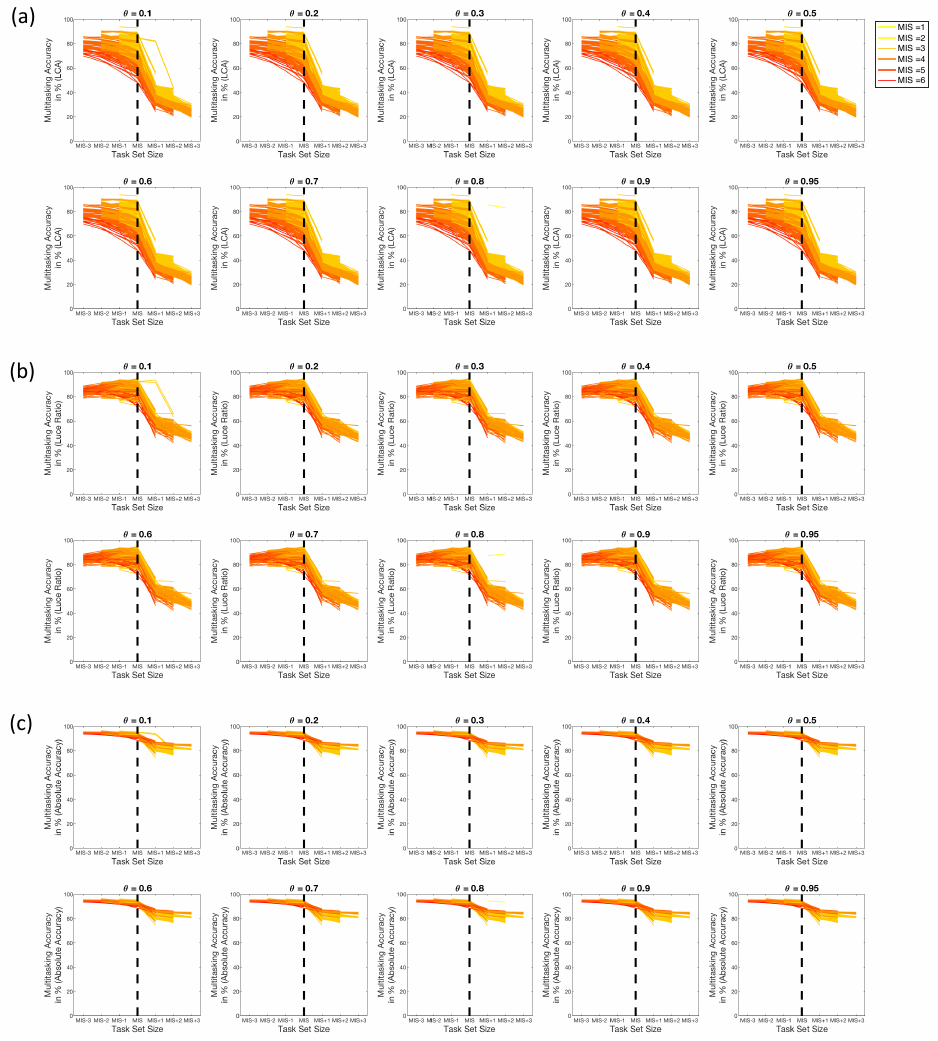}
\caption{\textbf{Multitasking accuracy of a network with 100 hidden units as a function of the predicted empirical MIS $\bar{\alpha}$}. The plot shows the highest multitasking accuracy of the network as a function of the number of tasks it is asked to perform in parallel (performance curve) as indicated in relation to the network's MIS. Each line corresponds to the multitasking performance of a trained network, whereas the color of each line indicates the MIS for that network that was extracted using different correlation thresholds. Multitasking accuracy of the network was assessed using (a) a leaky competitive accumulator, (b) the Luce ratio or (c) the absolute error of each task-relevant output dimension. }\label{fig:suppl_F1}
\end{figure}

\begin{figure}
\centering
\includegraphics[width=1\textwidth]{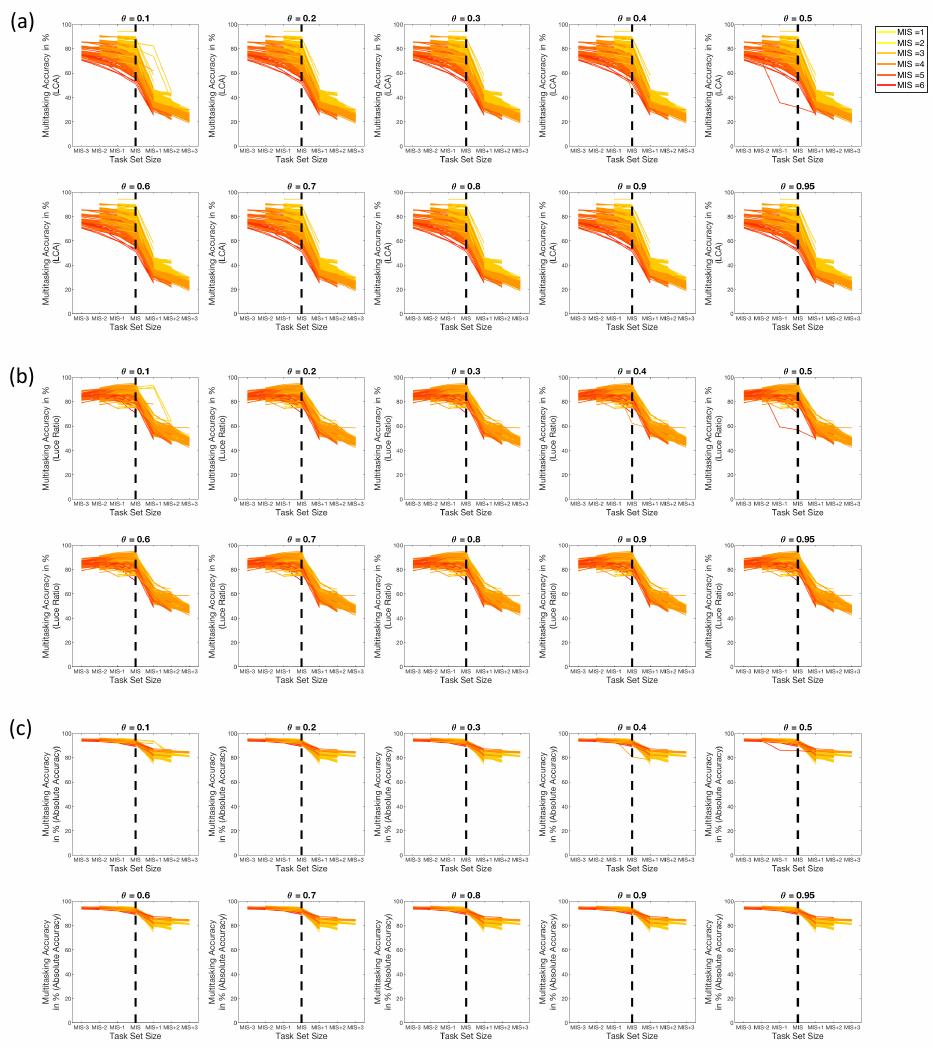}
\caption{\textbf{Multitasking accuracy of a network with 500 hidden units as a function of the predicted empirical MIS $\bar{\alpha}$.} The plot shows the highest multitasking accuracy of the network as a function of the number of tasks it is asked to perform in parallel (performance curve) as indicated in relation to the network's MIS. Each line corresponds to the multitasking performance of a trained network, whereas the color of each line indicates the MIS for that network that was extracted using different correlation thresholds. Multitasking accuracy of the network was assessed using (a) a leaky competitive accumulator, (b) the Luce ratio or (c) the absolute error of each task-relevant output dimension. }\label{fig:suppl_F2}
\end{figure}

\begin{figure}
\centering
\includegraphics[width=1\textwidth]{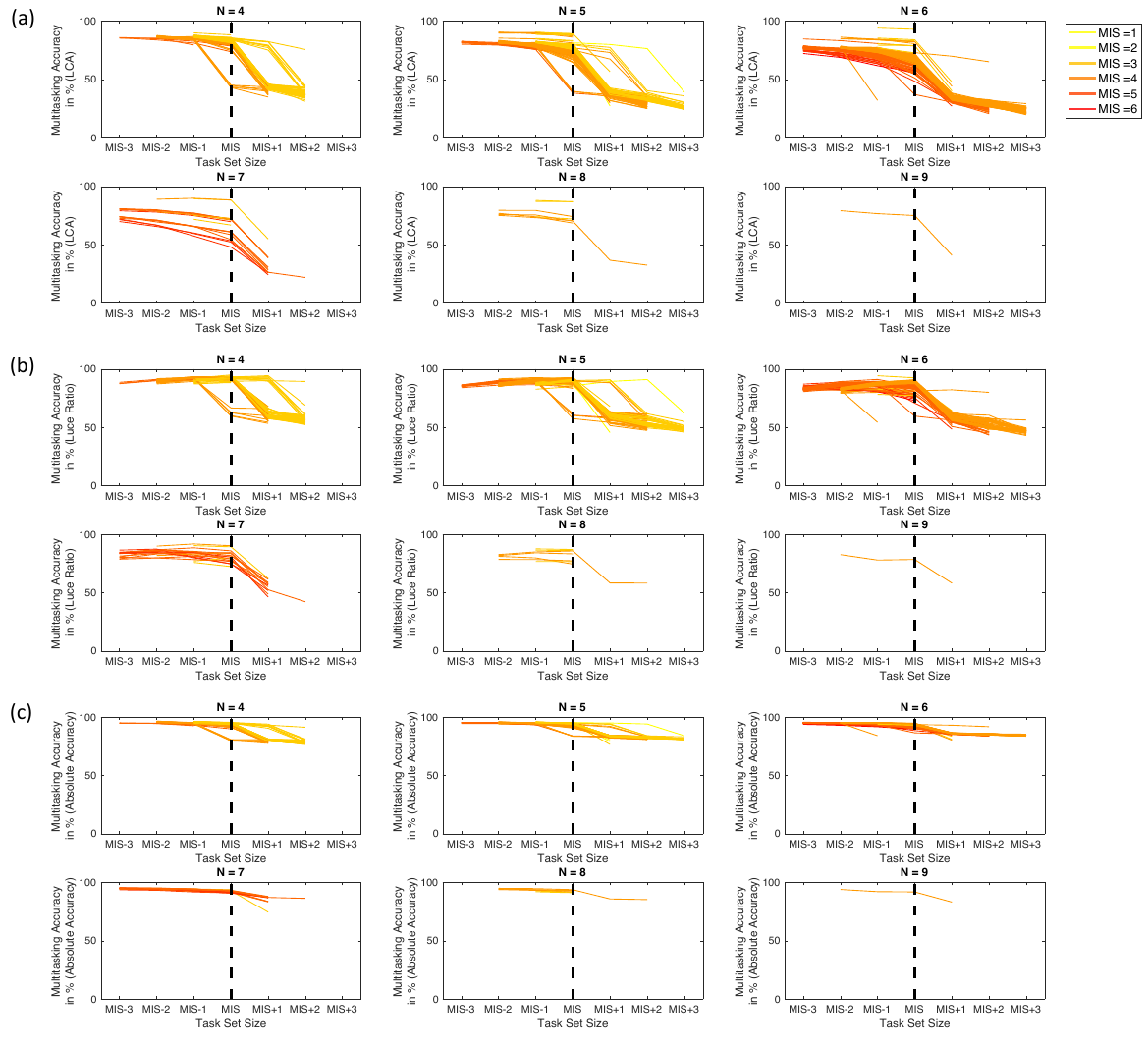}
\caption{\textbf{Multitasking accuracy of a network with 100 hidden units as a function of the predicted empirical MIS $\bar{\alpha}$}. Different columns correspond to a different number of input and output dimensions in the task environment. The plot shows the highest multitasking accuracy of the network as a function of the number of tasks it is asked to perform in parallel (performance curve) as indicated in relation to the network's MIS. Each line corresponds to the multitasking performance of a trained network for a given size (measured in terms of the number of input/output dimensions), whereas the color of each line indicates the MIS for that network that was extracted using different correlation thresholds. Multitasking accuracy of the network was assessed using (a) a leaky competitive accumulator, (b) the Luce ratio or (c) the absolute error of each task-relevant output dimension.}\label{fig:suppl_F3}
\end{figure}

We assessed the robustness of reported simulation results with respect to three parameters: (a) the correlation threshold $\theta$ that was used to decide whether two tasks share a representation in the trained network, (b) the metric that was used to assess multitasking performance of the network, and (c) the number of units in the hidden layer of the neural network. We first considered a neural network with 100 units in the hidden layer and extracted $\bar{\alpha}$ for different correlation thresholds, ranging from $0.1$ to $0.9$ in steps of $0.1$, as well as $0.95$. Moreover, we assessed multitasking performance for this network using different performance metrics, such as the probability of responding correctly using an LCA for a task-relevant output dimension (see previous section), the ratio between the activity of the correct output unit and the sum of all output units within a task-relevant output dimension (Luce ratio), as well as the absolute error between the correct response pattern and the actual response pattern in a task-relevant output dimension subtracted from 1. Figure \ref{fig:suppl_F1} shows the best multitasking accuracy across all task combinations of a given task set size (relative to the predicted empirical MIS $\bar{\alpha}$) across all parameter settings for a network with 100 hidden units. Figure \ref{fig:suppl_F2} shows the same set of results for a network with 500 hidden units. The results shown in both figures replicate the simulation results in the main text, namely that the best multitasking accuracy of the network drops as soon as the number of tasks in the set exceeds  $\bar{\alpha}$. Critically, we observed this pattern across a wide range of correlation thresholds, different performance metrics, as well as different number of hidden units in the network. 

Finally, Figure \ref{fig:suppl_F3} illustrates how the effects reported in Figure \ref{fig:suppl_F1} depend on the number of input and output dimensions available in the task environment. Simulation results indicate that the empirical MIS $\bar{\alpha}$ tends to overestimate the point at which the best multitasking performance drops as a function of task set size. Thus, $\bar{\alpha}$ serves as an upper bound for the multitasking capacity of the network as the complexity of the task environment increases. 

\clearpage
\section{Predicting Multitasking Accuracy for Networks with Random, Unconstrained Weights}

In the main text, we predict multitasking accuracy for networks with fixed weights from the task layer to the hidden layer. The weights were fixed to bias the learning of task representations either toward a basis set representation or a tensor product representation. Here, we assessed whether it is possible to predict multitasking accuracy based on learned representations of 400 networks (with 100 hidden units each) with randomly initialized, unconstrained weights. For each of the networks, we randomly initialized and trained all weights, including weights from the task layer to the hidden layer. We used the MIS predicted from the task environment, $\alpha$, as well as the MIS predicted from learned task representations, $\bar{\alpha}$ to predict multitasking accuracy as described in the Methods section of the main text. 

\begin{figure}[h!]
\centering
\includegraphics[width=1\textwidth]{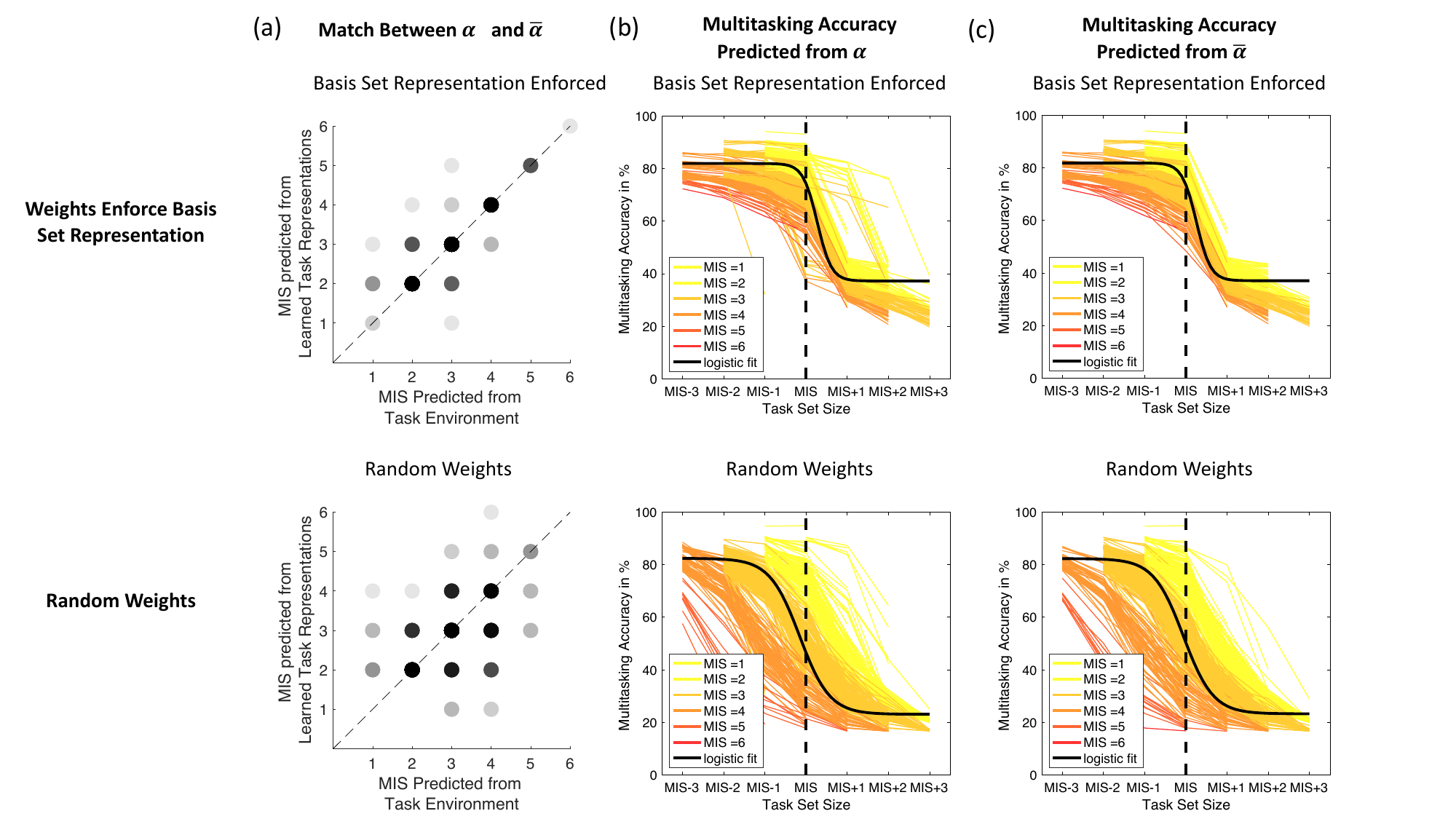}
\caption{\textbf{Simulation results for networks biased toward basis set representation (upper plots) and networks with randomly initialized, unconstrained weights (lower plots)}. (a) Match between the MIS predicted from the task structure graph, $\alpha$, and the MIS predicted from learned representations extracted from the neural network, $\bar{\alpha}$. The darker the color of a dot, the higher the proportion of networks representing this configuration. (b) Parallel processing capacity of the trained network is predicted by $\alpha$. (c) Parallel processing capacity of the trained network is predicted by $\bar{\alpha}$. Plots in (b, c) show the highest multitasking accuracy of the network as a function of the number of tasks it is asked to perform in parallel (performance curve) as indicated in relation to the network’s MIS. Each line corresponds to the multitasking performance of a trained network, whereas the color of each line indicates the predicted MIS for that network. The solid black line depicts the average fit of a logistic function to accuracy curves across networks. }
\label{fig:suppl_randomWeights}
\end{figure}

Figure \ref{fig:suppl_randomWeights} compares simulation results from 400 networks with fixed weights from the task layer to the hidden layer (biased toward basis set representation; see main text) against simulation results from randomly initialized, unconstrained networks. The latter networks show a greater mismatch between $\alpha$ and $\bar{\alpha}$, indicating that networks learned a wider spectrum of representations when weights from the task layer to the hidden layer were randomly initialized and trained. Both $\alpha$ and $\bar{\alpha}$ accurately predict maximum multitasking accuracy of randomly initialized, unconstrained networks, as the inflection point of the fitted performance curves for these networks lies significantly below a set size equal to $\alpha+1$, $t(328) = -27.6974, p < 10^{-87}$, and below a set size equal to $\bar{\alpha}+1$, $t(349) = -27.8839, p < 10^{-90}$. However, multitasking accuracy does not drop right after a set size equal to $\alpha$, $t(328) = -3.7042, p = 1.0
$, or after a set size equal to $\bar{\alpha}$, $t(349) = -1.5656, p = 0.94083$, indicating that both metrics overestimate multitasking capacity when all weights are randomly initialized and trained. Overall these results indicate that both $\alpha$ and $\bar{\alpha}$ can be used to predict an upper bound for multitasking accuracy in randomly initialized, unconstrained networks, although the bound is not as tight as the bound predicted for networks with an inductive bias toward a basis set representation.

\clearpage
\section{Predicting Multitasking Accuracy for Multi-Layer Networks}

To investigate how well multitasking metrics $\alpha$ and $\bar{\alpha}$ generalize to networks with more than one hidden layer, we simulated 290 networks with two subsequent hidden layers, each trained in a different task environment. Each of the two hidden layers consisted of 500 units and received weight projections from the task layer. To explore different forms of representations between environments, we fixed weight from the task layer to each hidden layer either to a basis set representation (tasks with the same input dimension are assigned the same set of units) or a tensor product representation (all tasks are assigned a different set of units). For each network and each hidden layer, we randomly assigned either of the two representations, each with a probability of $0.5$. Thus, a network could be biased to learn one of four possible network configurations: (1) basis set representation in both hidden layers, (2) basis set representation in the first hidden layer and tensor product representation in the second hidden layer, (3) tensor product representation in the first hidden layer and basis set representation in the second hidden layer, and (4) tensor product representation in both layers. All other weights were randomly initialized and trained as described in the main text. 

\begin{figure}[h!]
\centering
\includegraphics[width=0.6\textwidth]{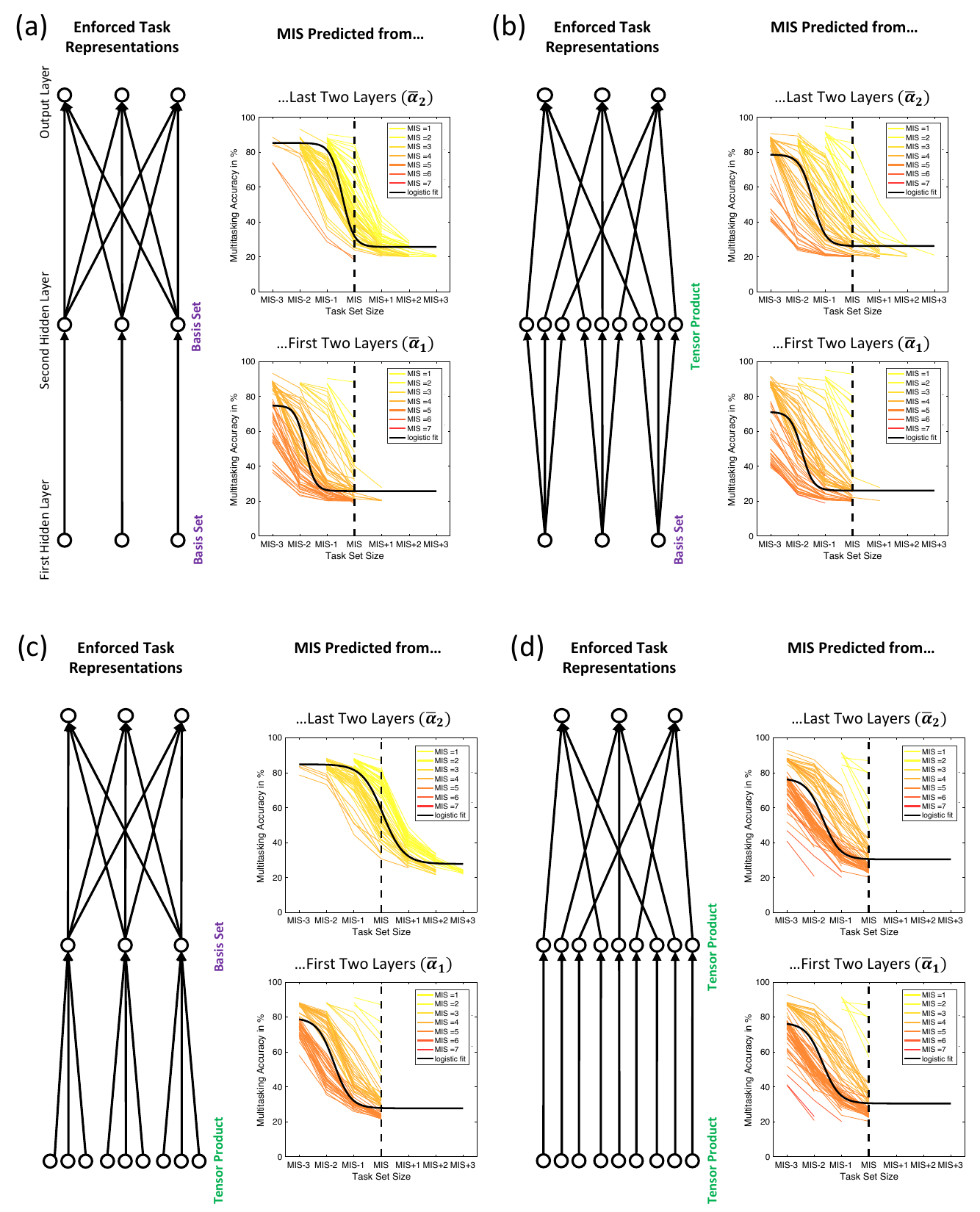}
\caption{\textbf{Simulation results for networks with two hidden layers.} The networks were trained with fixed weights from the task layer to each of the two hidden layers, enforcing either a basis set or a tensor product representation. Plots (a-d) show results for different enforced network configurations: (a) basis set representation in both hidden layers, (b) basis set representation in the first hidden layer, tensor product representation in the second hidden layer, (c) tensor product representation in the first hidden layer, basis set representation in the second hidden layer, (d) tensor product representation in both hidden layers. Each plot depicts a 2-layer task graph for a task environment with nine tasks (three input dimensions and three output representations). The first and second layers of the task graph depict representations enforced at the first and hidden layer of the neural network, respectively.  The last layer depicts the three output dimensions of the network. The two subplots in each panel show the highest multitasking accuracy of the network as a function of the number of tasks the network is asked to perform in parallel (performance curve) as indicated in relation to the network’s MIS, either predicted from learned representations at the first two layers, $\bar{\alpha}_1$ (lower subplot), or predicted from learned representations at the last two layers, $\bar{\alpha}_2$ (upper subplot). Each line corresponds to the multitasking performance of a trained network, whereas the color of each line indicates the predicted MIS for that network. The solid black line depicts the average fit of a logistic function to accuracy curves across networks. }
\label{fig:suppl_deepNet}
\end{figure}

After training, we extracted two task dependency graphs, one based on learned task representations in the first and second hidden layer, and one based on learned representations in the second hidden layer and in the output hidden layer, using the same method as described in the main text. We then computed the MIS for both dependency graphs, $\alpha_1$ for the first two layers and $\alpha_2$ for the last two layers, and used each to predict multitasking accuracy. Figure \ref{fig:suppl_deepNet} indicates simulation results, separated for each of the four network configurations. On average, we found that the MIS from from the first two layers and the last two layers both predict an upper bound for multitasking accuracy of trained networks, as the inflection point of the fitted performance curves lies significantly below a set size equal to $\bar{\alpha}_1+1$, $t(227) = -103.9067, p < 10^{-192}$, and below a set size equal to $\bar{\alpha}_2+1$, $t(234) = -31.9946,, p < 10^{-86}$. However, both $\bar{\alpha}_1$, $ t(227) = -66.3462, p = 1$, and $\bar{\alpha}_2$, $t(234) = -14.8443, p = 1.0$ overestimated the maximum number of tasks that a network can perform. To assess overall multitasking performance, we fitted a logistic curve to multitasking performance as a function of actual task set size (not normalized to the predicted MIS). We found that the inflection point for fitted curves is highest for networks with a tensor product representation enforced at both hidden layers ($M = 3.4148, SD \pm 0.3919$) or at the last hidden layer ($M = 3.326, SD \pm 0.2405$), followed by networks with a bias toward basis set representation in the first layer and tensor product representation in the last layer ($M = 2.7816, SD \pm 0.26816$) or a bias toward basis set representation in both layers ($M = 2.859, SD \pm 0.3017$). This suggests that tensor product representations can improve multitasking performance of the network, as long as subsequent layers maintain  such representational separation between tasks. 

\clearpage
\section{Empirical expected performance}
To test the formal analysis results on the effective parallel processing capacity, we derived measures of the empirical parallel processing capacity from 182 trained neural networks. For each trained network, we generated a task environment from a randomly sampled Erdos-Renyi graph $G_{TS}$ with a fixed degree density (0.2) and a given number of input and output nodes N where N was varied between 4 and 9 across different task environments. Each network was parameterized and trained on the task environment to criterion as described in the Methods section of the main text. We then computed the empirical effective parallel processing capacity as the expected reward for multitasking under a restrictive and permissive reward regime. To assess the expected reward for a restrictive reward regime, we measured the average proportion of multitasking combinations for a given task set size $p(\theta = \gamma)$ for which all tasks were performed with a maximum absolute error of 0.05. For example, if one would randomly pick three tasks (with the constraint that they are structurally independent) then this metric would correspond to the likelihood that the network would be able to perform all three tasks. The expected reward under such a permissive regime $\phi_\gamma$ corresponds to the number of attempted tasks multiplied by this likelihood (see Equation 5 in the main text). We computed the expected reward for a permissive reward regime $\tilde{\phi}_\gamma$ as the average number of tasks that the network would be able to perform with a maximum absolute error of 0.05 for a given task set size. For example, if one would pick three tasks (with the constraint that they are structurally independent) then this metric would correspond to the average number of tasks that the network could perform (ranging from zero to three). 

\begin{figure}[h!]
\centering
\includegraphics[width=.75\textwidth]{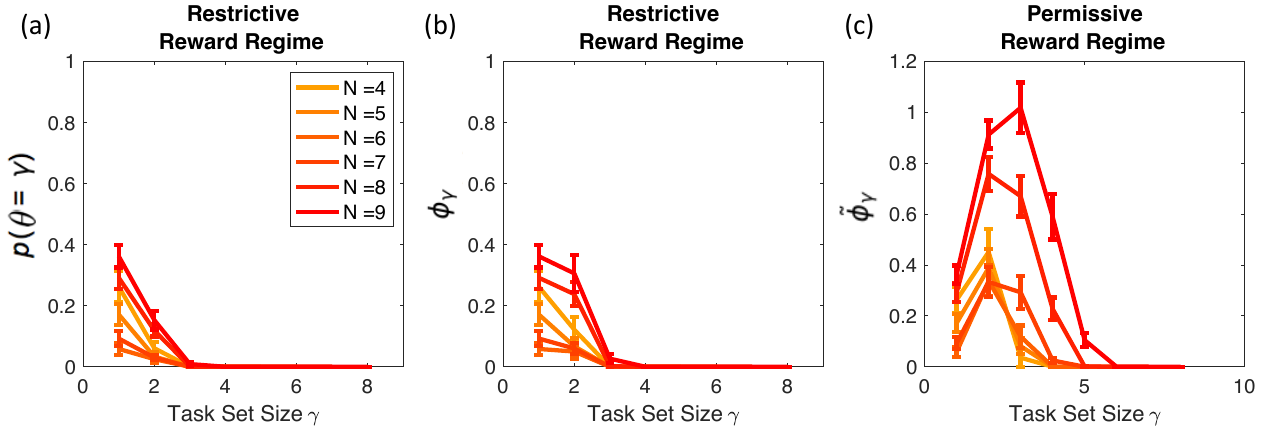}
\caption{\textbf{Empirical effective parallel processing capacity of trained neural networks for restrictive and permissive reward regimes.} (a) The average proportion of multitasking combinations for which all tasks were performed with a maximum absolute error of 0.05 probability as a function of task set size $\gamma$. (b) Expected reward under restrictive reward regime, measured as the number of attempted tasks multiplied by the likelihood shown in (a). (c) Expected reward under permissive reward regime, measured as the average number of tasks that a network would be able to perform with a maximum absolute error of 0.05 as a function of task set size $\gamma$. These results are in qualitative agreement with the theoretical results in Section \ref{sec:theorretical_expected_performance}, Figure \ref{fig:ER-fixed-density-ts.pdf}.
}\label{fig:suppl_F4}
\end{figure}

Our simulation results (Figure \ref{fig:suppl_F4}) show qualitative similarities with the formal analysis results (Figure \ref{fig:ER-fixed-density-ts.pdf}). 
First, we observed that $p(\theta = \gamma)$  as well as $\phi_\gamma$ converge to zero as the set 
size $\gamma$ increases (more details in Section \ref{sec:theorretical_expected_performance}). 
Unlike the analytic prediction, the expected reward for a permissive reward regime $\tilde{\phi}_\gamma$ does also converge to zero, reflecting an even more restrictive limitation on the effective parallel processing capacity, possibly due to unaccounted interference between tasks in trained artificial networks. 

\clearpage
\section{Computation of the MIS of the dependency graph}
The Maximum Independent Set (MIS) problem \cite{Tarjan_SJC_77} is a particular instance of a larger optimization problem class, called Maximum Packing Set (MPS) problem \cite{hazan2006complexity}, which we introduce below. 
Given a set $ = \{1,\ldots,\}$ and a collection of its subsets $\mathrm{S} = \{S_i | S_i \subset ,  \in \}$ labeled by $ = \{1,\ldots,\}$.
A set packing is a collection of the pairwise disjoint subsets $S_i$, its size is the packing number.
The problem of finding the maximum packing number can be formulated as an integer programming problem as follows:
    \begin{align}
    \text{maximise} \quad \sum_{i\in V} n_i, \\
    \text{subject to } \sum_{i:r\in S_i} n_i \leq 1 \quad \forall r \in F\\
    \qquad n_i = \{0,1\} \forall i \in V
    \end{align}
We follow here the approach of \citet{Lucibello:2014ga}. 
Given a variable nodes set $V$, we assign to each $i \in V$ a variable $n_i$ that takes values in  $\{0, 1\}$. 
The factor nodes set $F$ contains instead the elements acting as constraints on the variables $n_i$. The edge set $E$ is then defined as $E = \{(i,r) | i\in V, r \in S_i \subset F\}$, specifying connections between variables and factors.

Using the factor graph G composed by these three sets,  $G=(V, F, E)$, the MSP problem specification can be rewritten as
\begin{align}
\sum_{i\in\partial_r} n_i \leq 1 \quad \forall r \in F
\end{align}
where $\partial_r$ is the neighbourhood of node $r$ in G, corresponding to the set of variable nodes that are involved in factor $r$. 
Hence, a solution is then given by configuration of $n_i$ values that satisfy the condition above. 

It is possible to write analytical expressions for the expected MIS density, $\rho({n_i}) = \frac{1}{N} \sum_{i\in V} n_i $, where $N = |V|$. 
These expressions are exact in the sparse regime, effectively amounting to local tree-likeness of the factor graph. 
The expressions for Replica Symmetry solution are \cite{Lucibello:2014ga}:
\begin{align}
\rho^{RS} = \frac{\langle k \rangle }{\langle c \rangle} \left( 1 - \mathbb{E}_c \left(1 - \mathbb{E}_{\tilde{k}} p_*^{\tilde{k}} \right)^{\tilde{c}} \right) + \mathbb{E}_{k} (1-k)p_*^{k} \label{eq_rho}\\
\text{where $p_*$ satisfies} \quad p_*  = \mathbb{E_{\tilde{c}}}\left( 1 - \mathbb{E}_{\tilde{k}} p_*^{\tilde{k}} \right)^{\tilde{c}} \label{eq_pstar}
\end{align}
where $c$ is the degree of a factor node, $\tilde{c}$ the excess degree of a factor node, $k$ the degree of a variable node, $\tilde{k}$ is the excess degree of a factor node, and $\langle c \rangle$ and $\langle k \rangle $  the average factor and variable degrees respectively. 
The expectations over the various degrees listed above are denoted by the corresponding $\mathbb{E}_{\tilde{c},\tilde{k},c,k}$.  
In standard uncorrelated networks the excess degree distributions take the form:
$P (c = s) = \frac{(s+1) P(c = s+1)}{\langle c \rangle}$ and $P (k = s) = \frac{(s+1) P(k = s+1)}{\langle k \rangle }$. 
Exploiting the properties of the degree and excess degree distributions, they can be rewritten in terms of generating functions as follows:
\begin{align}
\mathbb{E}_{\tilde{k}} {p_*}^{\tilde{k}} =   
\sum_{\tilde{k}} p_*^{\tilde{k}} p(\tilde{k}=s) =  
\frac{\sum_{s=0}^\infty e^{(\ln p_*) s} (s+1) p(s+1)}{\langle k \rangle } = 
\frac{\sum_{s=0}^\infty e^{t s} (s+1) p(s+1)}{\langle k \rangle } \nonumber \\ 
\frac{1}{p_*}  \sum_{s=0}^\infty \frac{e^{t (s+1)} p(s+1) (s+1)} {\langle k \rangle } =
\frac{1}{\langle k \rangle  p_*}  \sum_{s=0}^\infty \frac{d}{dt} e^{t (s+1)} p(s+1)  = \nonumber \\
\frac{1}{\langle k \rangle  p_*}  \frac{d}{dt} \left( M_k(t) -1  \right)  =  \frac{1}{\langle k \rangle  p_*}  M_k'(t)
\end{align}
where $t=\ln p_*$ and $M_k(t)$ is the moment generating function of the degree distribution and the derivative of $M_k$ is taken over $t$. Putting this expression back into Eq. \ref{eq_pstar}, we finally obtain:
\begin{equation}
p_*  = \mathbb{E}_{\tilde{c}}\left( 1 - \frac{1}{\langle k \rangle  p_*} M_k'(t) \right)^{\tilde{c}}
\end{equation}

\begin{figure}
\centering
\includegraphics[width=\textwidth]{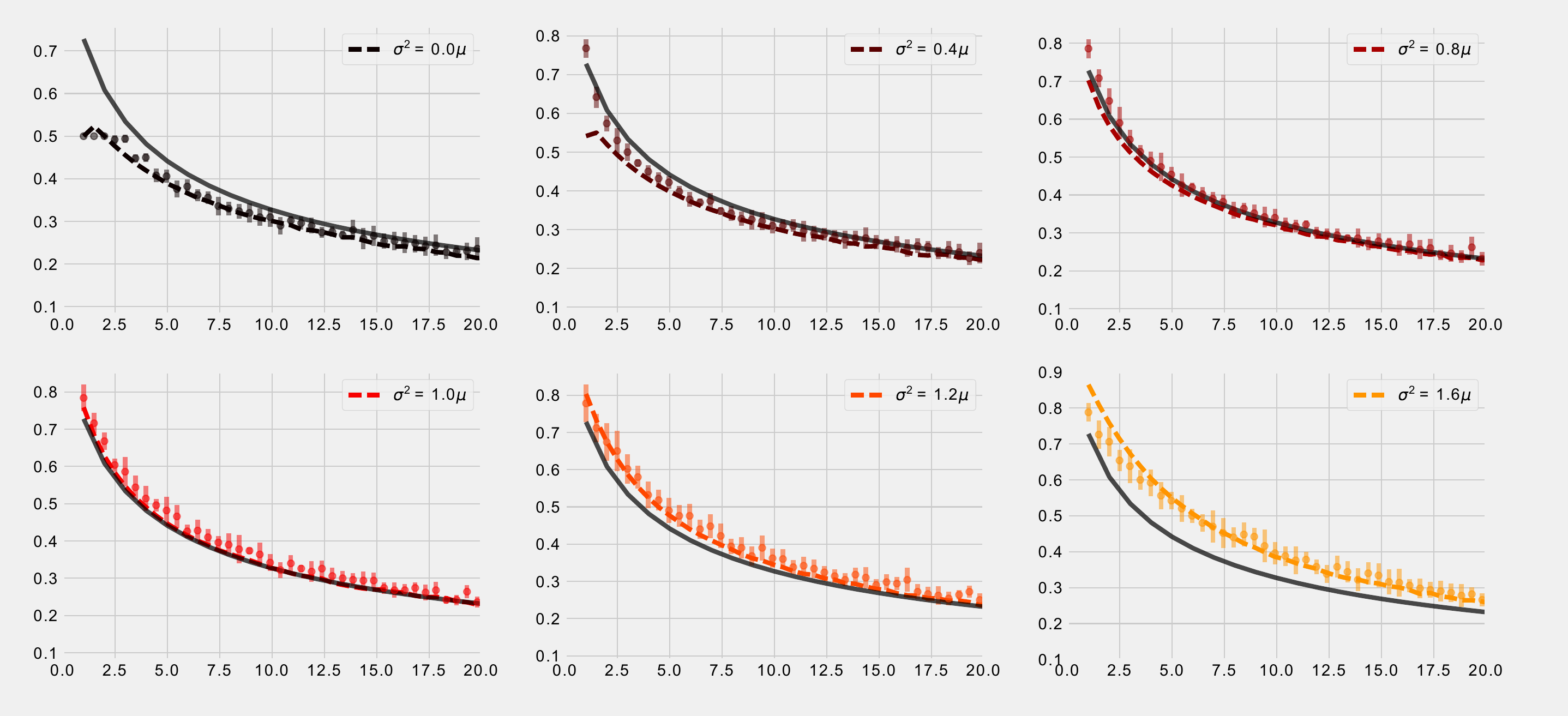}
\caption{\textbf{Effects of network heterogeneity on predicted $\rho_\alpha$.} The subplots shows $\rho_\alpha$ as a function of network average degree $\langle k \rangle $ for a range of network degree heterogeneities. We consider here a Gaussian degree distribution with average $\mu \in (0,20)$ and variance $\sigma^2$ proportional to $\mu$ as shown in the legends. The solid black line is the prediction for a Poissonian graph, while the dashed line is the corresponding analytical solution of Equation \ref{eq:gaussian_solution}. We see very clearly that for larger $\sigma^2$, $\rho_\alpha$ systematically shifts to higher values, staying below the Poissonian solution for $\sigma^2<1$ and above it for $\sigma^2 > 1$. }
\label{fig::gaussian_heterogeneity}
\end{figure}

The second term of Eq.~\ref{eq_rho} can be written as follows:
\begin{equation}
\mathbb{E}_{k} [(1 - k) p_*^{k}]  = 
\mathbb{E}_{k} p_*^{k} - \mathbb{E}_{k} [k p_*^{k}] \, 
\end{equation}
The first term of the equation becomes:
\begin{equation}
\mathbb{E}_{k} p_*^{k} = \sum_k p(k) p_*^k = \sum_k p(k) e^{(\ln p_*)k} = \sum_k p(k) e^{tk} = M_k(t)
\end{equation}
where again $t = \ln p$, and $M(t)$ is the generating function for the network's degree distribution.\\
The second term becomes instead:
\begin{equation}
\mathbb{E}_{k} [k p_*^{k}] = \sum_k p(k) k p_*^k = \sum_k p(k) k e^{(\ln p_*) k} = \sum_k p(k) \frac{d}{dt} e^{tk} = \frac{d}{dt} \sum_k p(k) e^{tk}  = M_k'(t)
\end{equation}
Collecting the two terms, we obtain:
\begin{equation}
 \mathbb{E}_{k} [(1 - k) p_*^{k}]  =   M_k(t) - M_k'(t)
\end{equation}
\noindent
and Eqs. \ref{eq_rho} and \ref{eq_pstar} can be rewritten as:
\begin{eqnarray}\label{eq:general_solution}
p_*  = \mathbb{E}_{\tilde{c}}\left( 1 - \frac{1}{\langle k \rangle  p_*} M_k'(t) \right)^{\tilde{c}} \\ 
\rho^{RS} = \frac{\langle k \rangle }{\langle c \rangle} \left( 1 - p_*^{c/(c-1)} \right) + M_k(t) - M_k'(t)
\end{eqnarray}
These allow us to plug in the moment-generating functions corresponding to the degree distribution of interest. 
In the main text, we use a slightly different notation. 
This is due to the fact that for the case of MIS $c = 2$ for all factor nodes, so we can simplify the expressions. \\
For a generic Gaussian-like distribution,  the moment generating function takes the form $M_k(t) = e^{\langle k \rangle  t + \sigma^2 t^2/2}$. Using the expression above, we obtain:
\begin{equation}
p_* = \left [1 - \frac{1}{\langle k \rangle  p_*} (\langle k \rangle  + \sigma^2 \ln p_*) M_k(\ln p_*) \right]\\
\end{equation}
and 
\begin{equation}\label{eq:gaussian_solution}
\rho_\alpha = \frac{\langle k \rangle }{\langle c \rangle} \left( 1 - p_*^{c/(c-1)} \right) + M_k(\ln p_*) (1 - \langle k \rangle  - \sigma^2 \ln p_*)
\end{equation}
While for a Poisson degree distribution, the generating function for Poisson distribution $M_k(t) = e^{\nu (e^t-1)}$ and in our case with $\nu=\langle k \rangle , t=\ln p_*$, yielding:
\begin{align}
p_* = \left(1 - e^{\langle k \rangle (p_*-1)} \right)^{c-1} \text{yielding} \\
\rho = \frac{\langle k \rangle }{\langle c \rangle} \left(1 - p_*^{c/(c-1)} \right) + (1-p_* \langle k \rangle )e^{\langle k \rangle (p_*-1)}
\end{align}
In Figure \ref{fig::gaussian_heterogeneity}  we show that this expression captures well the behavior of $\rho_\alpha$ for increasing network density and for various levels of degree heterogeneity.   
Moreover, it gives an analytical grounding to the previous empirical observation that a larger heterogeneity of task overlap at fixed density results in a higher $\rho_\alpha$ \cite{SourceCode,Feng_et_al_2014}. 

\begin{figure}
\centering
\includegraphics[width=0.45\textwidth]{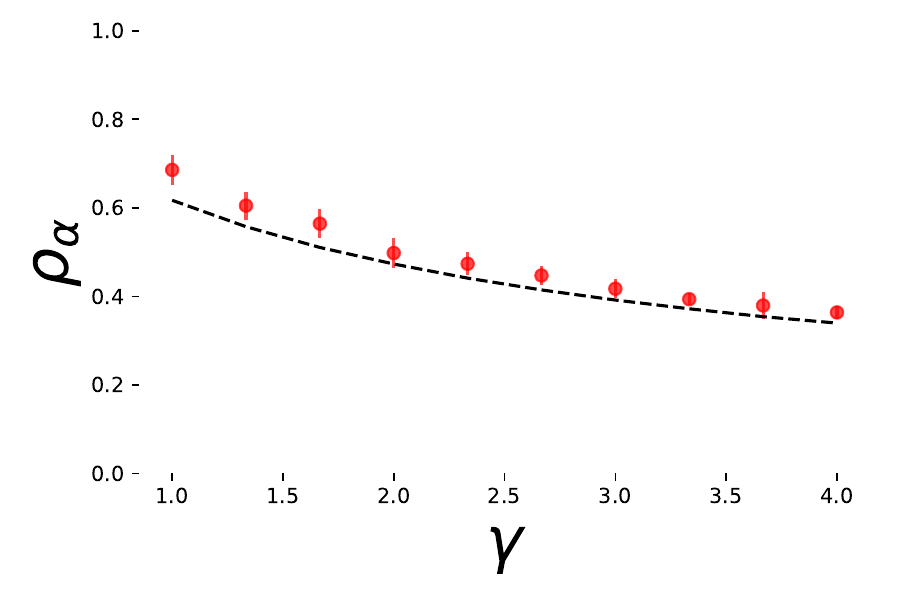}
\includegraphics[width=0.45\textwidth]{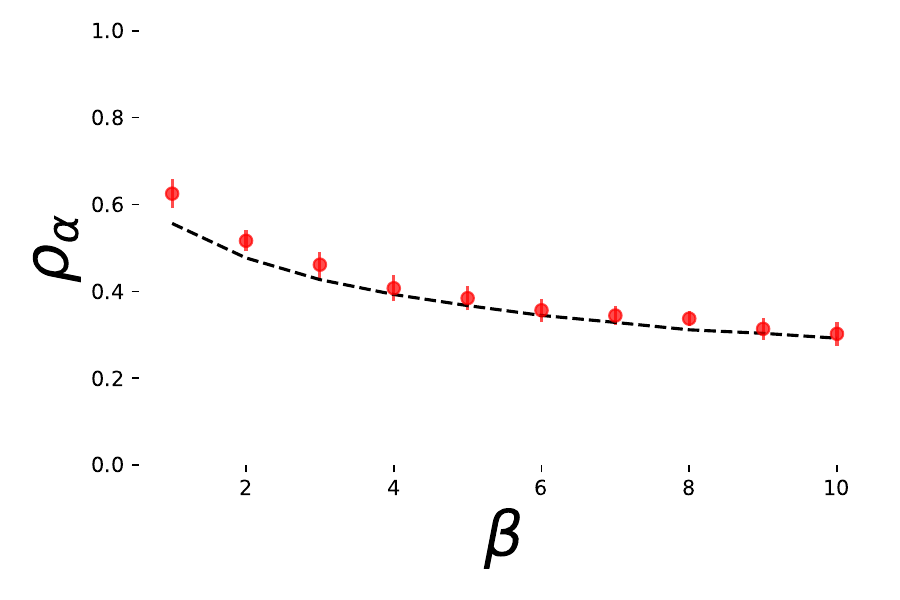}
\caption{\textbf{$\rho_\alpha$ results for Gamma degree distributions.}
 We show here the comparison of the measured  (dots) versus the predicted (dashed lines) $\rho_\alpha$ for random graphs with $N=70$ nodes and prescribed Gamma degree distribution $p(k) = \frac{1}{\Gamma[\gamma]\theta^\gamma} k^{\gamma} e^{-k/\theta}$ with varying  $\gamma$ of the degree distribution (\textit{left}, with fixed $\beta=3$), or  $\beta$ (\textit{right}, with fixed $\gamma=1$. For each parameter choice, we created 50 graphs.}
 \label{fig:gamma}
\end{figure}

\begin{figure}
\centering
\includegraphics[width=0.45\textwidth]{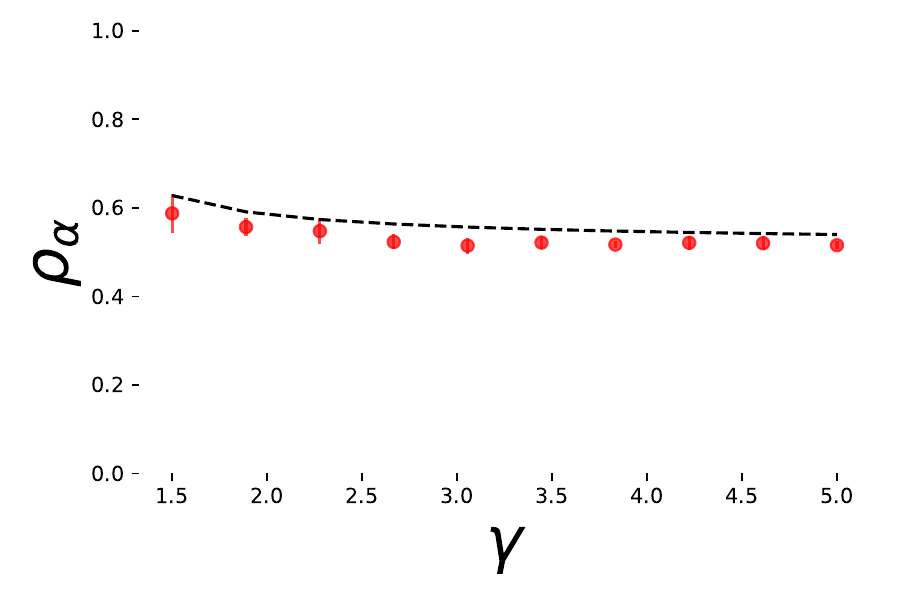}
\includegraphics[width=0.45\textwidth]{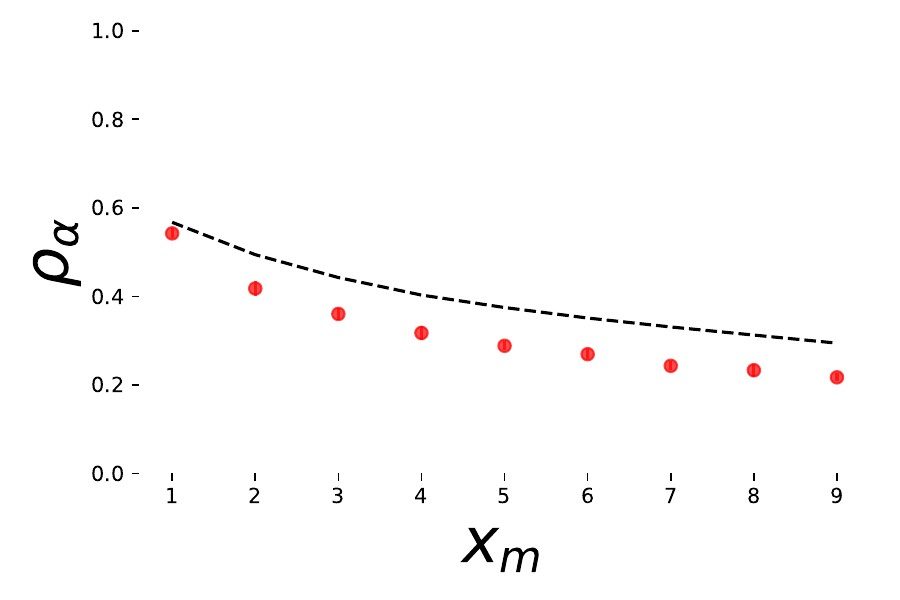}
\caption{\textbf{$\rho_\alpha$ results for Pareto degree distributions.} We show here the comparison of the measured  (dots) versus the predicted (dashed lines) $\rho_\alpha$ for random graphs with $N=200$ nodes and prescribed Pareto degree distribution $p(k) = \frac{\gamma x_m^\gamma}{k^{\gamma +1}}$ varying the slope $\gamma$ of the degree distribution (\textit{left}, with fixed $x_m=1$) and the scale of the minimum degree $x_m$ (\textit{right}, with fixed $\gamma=2.5$). For each parameter choice, we created 50 graphs.
For a Pareto degree distribution, the average degree is $\gamma x_m^\gamma / (\gamma -1 )$ so increasing $x_m$ directly increases the average degree, in turn decreasing in $\rho_\alpha$. 
s}\label{fig:pareto}
\end{figure}

In the manuscript we focused on normal and Poisson degree distributions for simplicity of explanation and for consistency with previous work by Feng et al. 
However, Eq. \ref{eq:general_solution} gives good results also with more structured degree distributions. 
We give here the solutions for Gamma and Pareto distribution.\\
For a Gamma degree distribution: 
$p(k) = \frac{1}{\Gamma[\gamma]\theta^\gamma} k^{\gamma} e^{-k/\theta}$ we obtain:
\begin{align}
p_* = 1 - \frac{1}{p_*(1-\theta\log p_*)} \left( \frac{1}{1- \theta \log p_*} \right)^{\gamma} \\
\rho = \frac{\langle k \rangle }{\langle c \rangle} \left(1 - p_*^{c/(c-1)} \right)  
+ \left(1 - \frac{\gamma \theta}{1-\theta \log p_*}\right) \left(\frac{\gamma \theta}{1-\theta \log p_*}\right)^{\gamma}
\end{align}
For a Pareto degree distribution $p(k) = \frac{\gamma x_m^\gamma}{k^{\gamma +1}} $, the density expression becomes:
\begin{align}
p_* = \left(1 - \frac{1}{\langle k \rangle p_*} (-\gamma^2 x_m^\gamma \Gamma[-\gamma, x_m \log p_*] (-\log p_*)^{\gamma-1} - \frac{\gamma}{\log p_*} p_*^{x_m} )   \right) \\
\rho = \frac{\langle k \rangle }{\langle c \rangle} \left(1 - p_*^{c/(c-1)} \right) 
+ \gamma x_m^\gamma \Gamma[-\gamma,-x_m \log p_*] \left( (-t)^\gamma + \gamma(-\log p_*)^{\gamma -1} \right) + \frac{\gamma}{\log p_*}p_*^{x_m}
\end{align}
\noindent In Figures \ref{fig:gamma} and \ref{fig:pareto} we show the comparison of simulated and predicted $\rho_\alpha$ for these two degree distributions. 

\clearpage
\section{Computation of the expected performance}
\label{sec:theorretical_expected_performance}
When we calculate the maximum parallel capacity $\alpha$ for a certain task graph $G_D$ with task-set (i.e. nodeset) $T$, we focus only on one (or a few, in some cases) maximum independence sets. It is however important to ask a different question: given a certain task set $T'\subset T$ with cardinality $\gamma = |T'| \leq \alpha$, what is the average number of tasks that can be performed in parallel? 
Both reward schemes,  $\phi_\gamma$ and $\tilde{\phi}_\gamma$, introduced in the main text, depend on the probability of finding independent task subsets $p(\theta, \gamma, G^D)$. 

\begin{figure}[h!]
\centering
\includegraphics[width=0.7\textwidth]{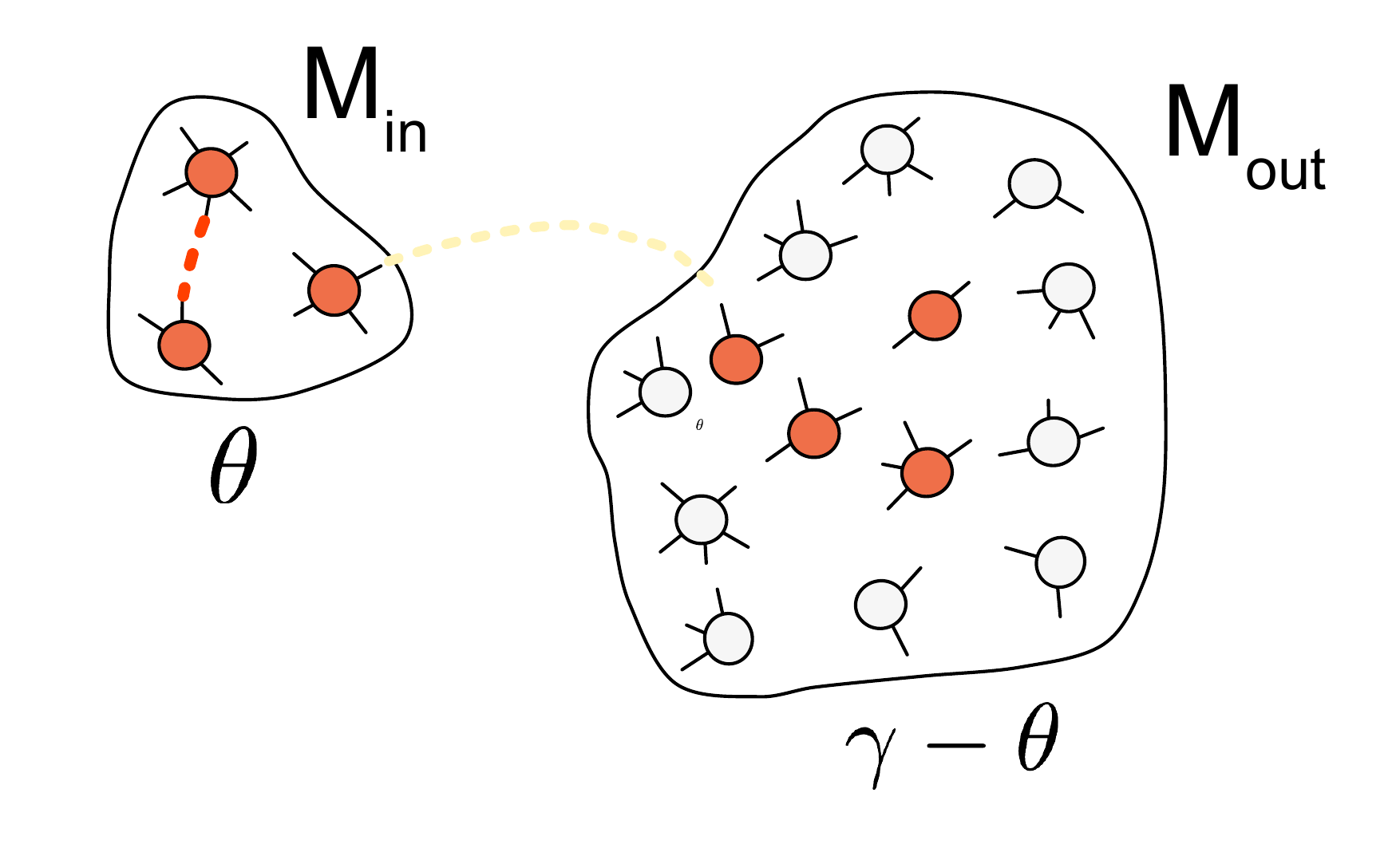}
\caption{\label{fig:av} Cartoon of the necessary conditions for $\theta$ (3 in this case) tasks $t_0, t_1, \ldots, t_{\theta-1}$ belonging to a subset $T'\subset T$ with cardinality $|T'|=\gamma$ ($\gamma=7$ in the example) of the entire task universe $T$ to be independent.}\label{fig:cartoon}
\end{figure}

\begin{figure}[h!]
\centering
\includegraphics[width=0.75\textwidth]{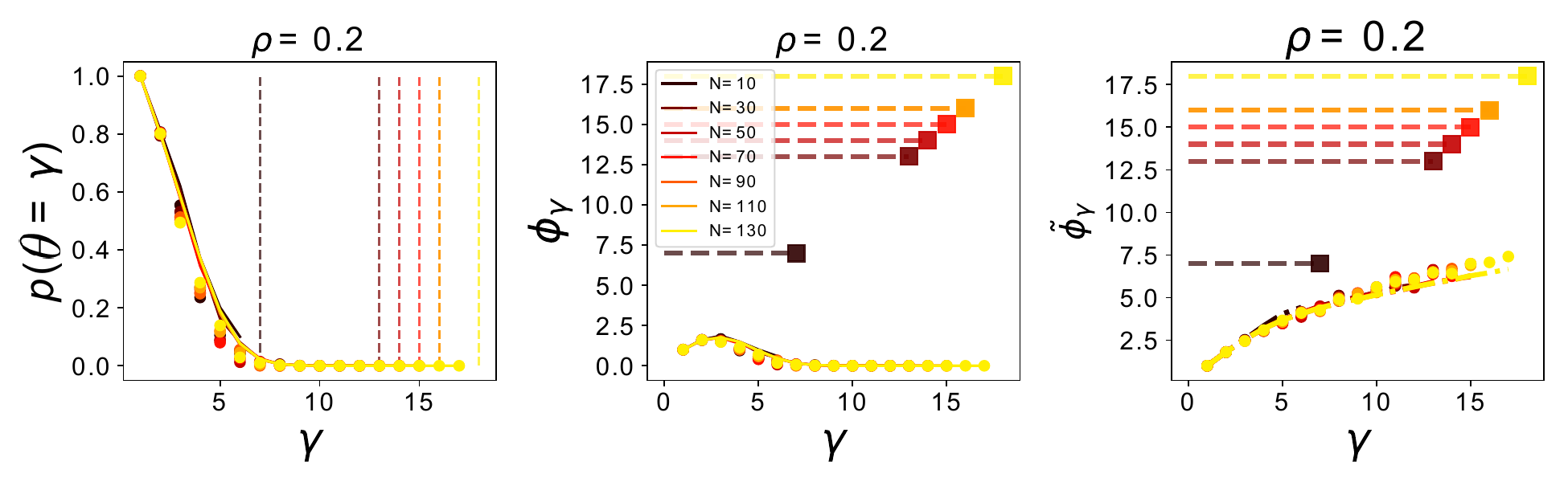}
\includegraphics[width=0.75\textwidth]{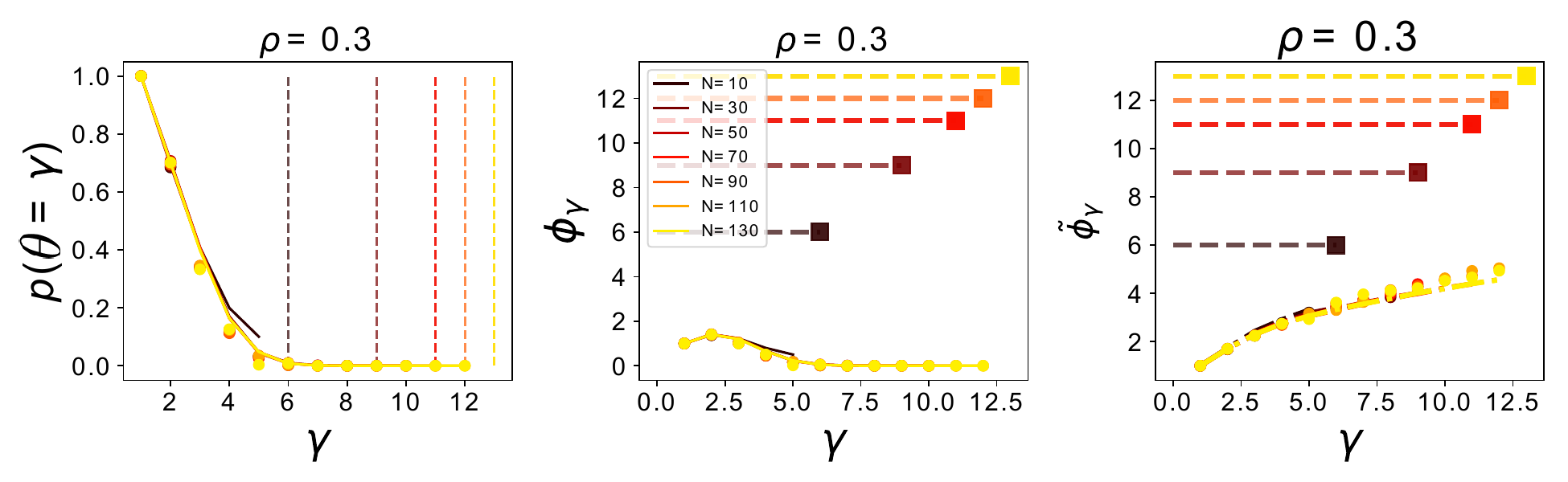}
\includegraphics[width=0.75\textwidth]{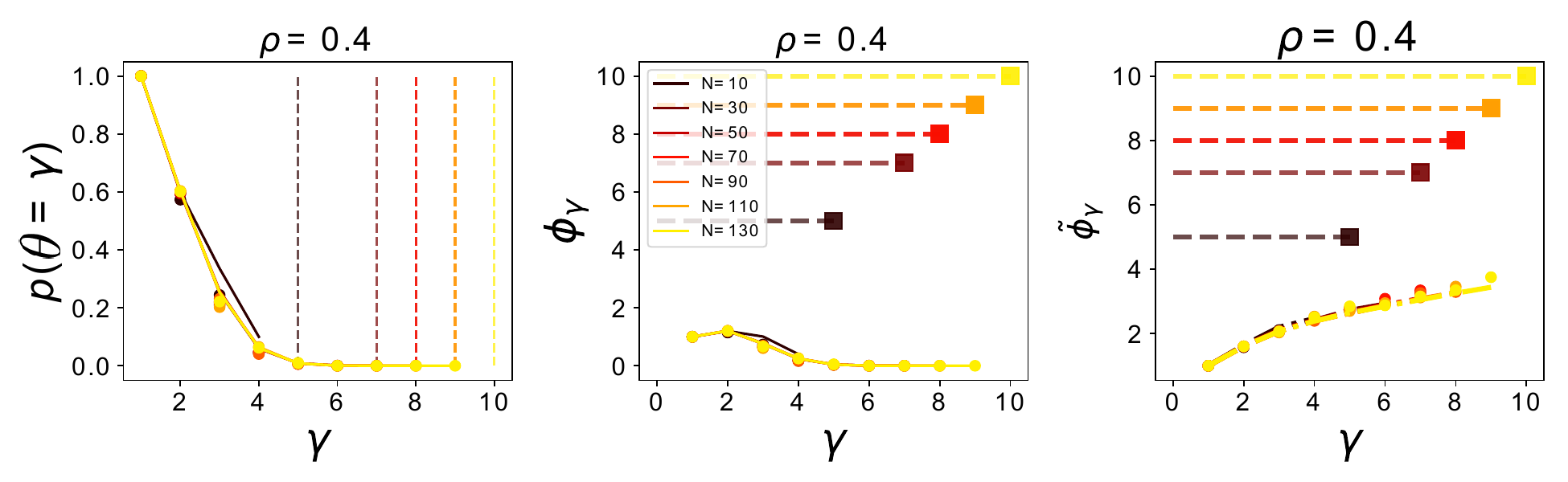}
\includegraphics[width=0.75\textwidth]{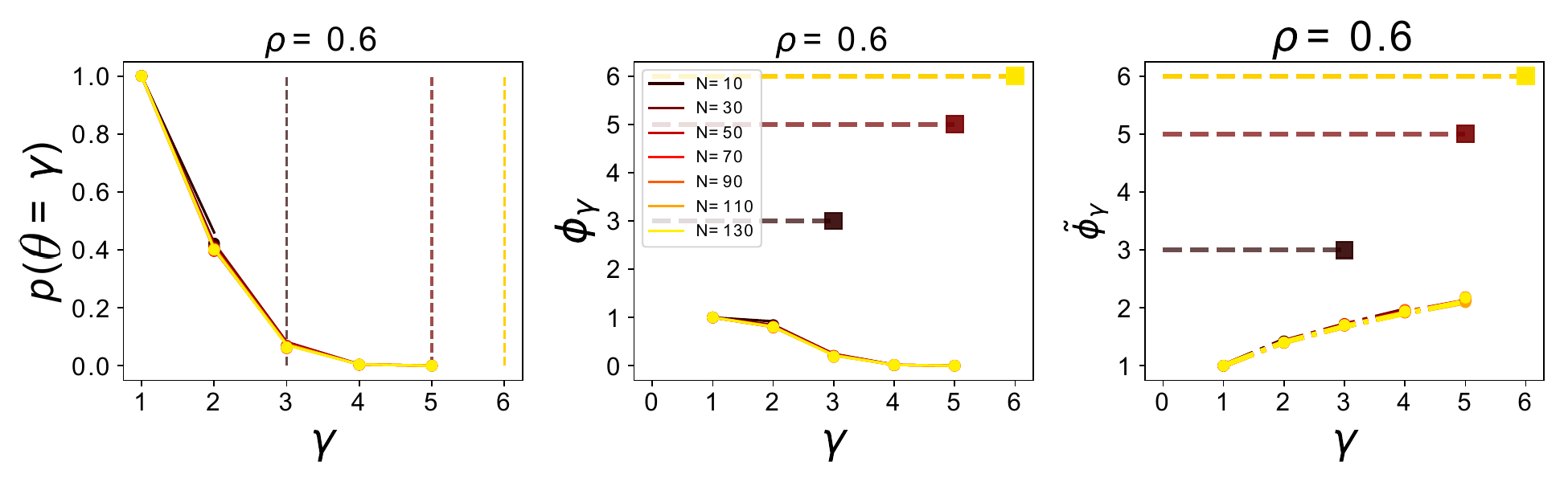}
\caption{\textbf{$p_\gamma$, $\phi_\gamma$ and $\tilde{\phi}_\gamma$ for a range of density of $G_D$.} Results from simulations (dots) for multiple sizes and density have good agreement with the analytical approximation (solid lines). Dashed vertical (left) and horizontal (center and right) lines represent the predicted MIS values for the corresponding size and density.}\label{fig:ER-ig-densities.pdf}
\end{figure}

\begin{figure}[h!]
\centering
\includegraphics[width=0.75\textwidth]{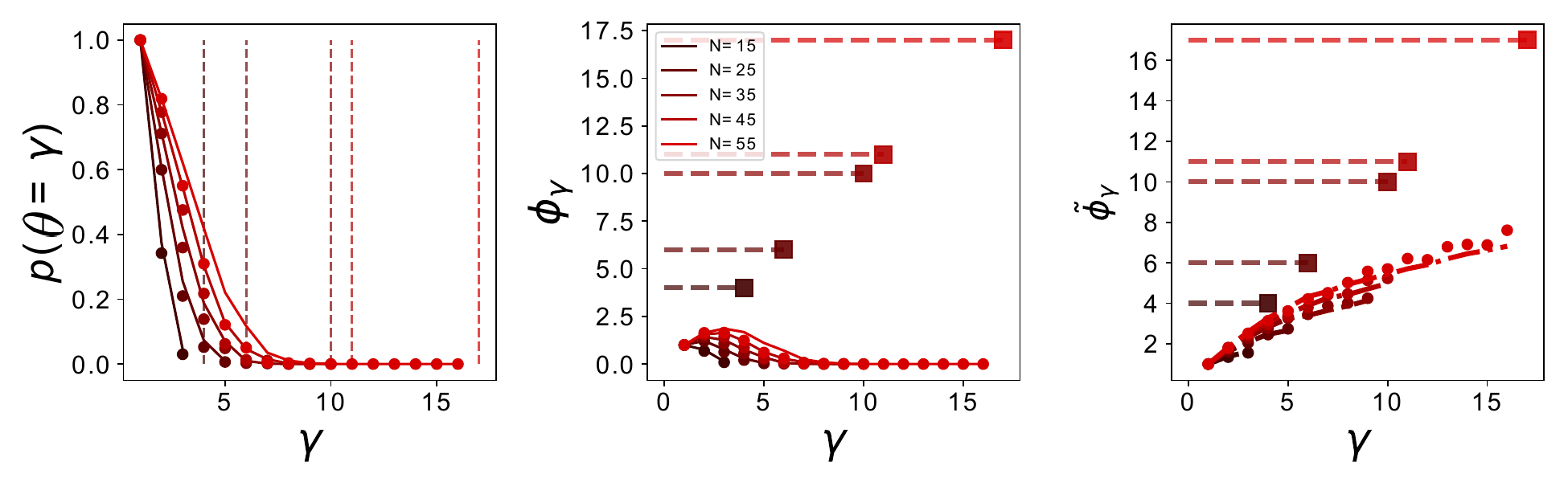}
\caption{\textbf{Results for $p_\gamma$, $\phi_\gamma$ and $\tilde{\phi}_\gamma$ for a range of network sizes at fixed average degree.} While we observe a small size dependence, results from simulations (dots) for multiple sizes have good agreement with the analytical approximation (solid lines) and display the same behaviour observed in Figure \ref{fig:ER-ig-densities.pdf}. Dashed vertical (left) and horizontal (center and right) lines represent the predicted MIS values for the corresponding size and density.}\label{fig:ER-fixed-degree.pdf}
\end{figure}

\begin{figure}[h!]
\centering
\includegraphics[width=0.75\textwidth]{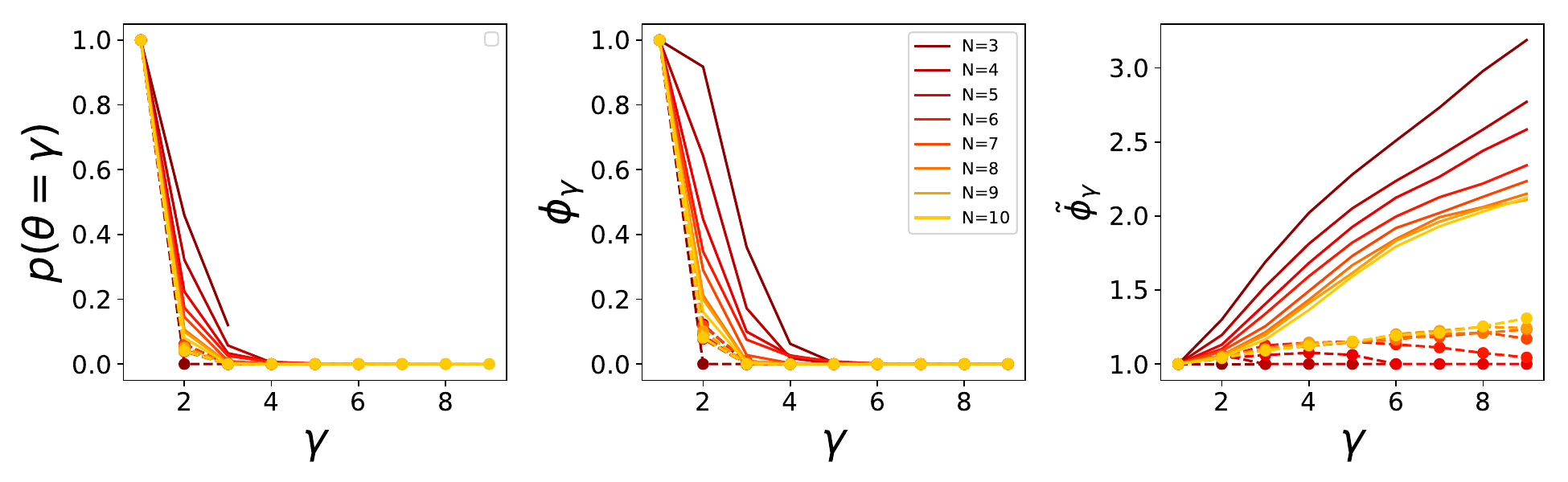}
\caption{\textbf{Numerical results for small $p_\gamma$, $\phi_\gamma$ and $\tilde{\phi}_\gamma$ for small $N$ with fixed density of the task structure graph.} Results from simulations (dots) for multiple sizes have good agreement with the analytical approximation (solid lines) and display similar qualitative behaviour observed in Figure \ref{fig:suppl_F4}.}
\label{fig:ER-fixed-density-ts.pdf}
\end{figure}

Below we give an estimate of this quantity using the degree distribution of $G_D$.
Denote the number of nodes in $G_D$ as $M$ and the number of edges in $G_D$ as $M_D$. 
The computation of $p(\theta, \gamma, G^D)$ can be easily visualized as in Figure \ref{fig:av}: 
denote the degree of the $\gamma$ nodes as degrees $\{k_0, \ldots, k_{\gamma-1}\}$; 
i) we need the $\theta$ nodes to be independent, that is, we are prohibiting the red edge in Figure \ref{fig:av}, which for each pair of nodes happens with probability $(1 - \frac{k_i k_j}{2M_D})$; 
ii) we then require the remaining $\gamma - \theta$ nodes to connect to at least one of the $M_{in}=\sum_{i=0}^{\theta-1} k_i$ stubs belonging to the independent $\theta$ nodes, and this happens with probability $M_{in} k_l /2M_D $ per node (we impose the existence of at least one yellow edge). 
Clearly $M_{out} = M_D - M_{in}$. 
Putting these contributions together we finally arrive at the probability of executing successfully $\theta$ out of $\gamma$ tasks with degrees $\{k_0, \ldots, k_{\gamma-1}\}$:  
\begin{equation} \label{prob:beta:gamma}
p(\theta;\gamma, G^D) \propto \left[ \prod_{i<j, i =0}^{\theta -1} \left(1 - \frac{k_i k_j}{2M_D} \right) \right] \left[\prod_{l=0}^{\gamma - \theta - 1} k_l\right] 
\left(\frac{M_{in}}{2M_D}\right)^{\gamma - \theta}
\end{equation}
Naturally, the full probability should include the probability of the degree configuration $p(\{k_0, \ldots, k_{\gamma-1}\})$ which for uncorrelated random graphs factorises in the product of $\gamma$ times $G_D$'s degree distribution. 
So, the expression for $p(\theta;\gamma, G^D)$ finally becomes:
\begin{equation}
p(\theta;\gamma, G^D) =  \left[ \prod_{i<j, i =0}^{\theta -1} \left(1 - \frac{k_i k_j}{2M_D} \right) \right] \left[\prod_{l=0}^{\gamma - \theta - 1} k_l\right]
\left(\frac{M_{in}}{2M_D}\right)^{\gamma - \theta} [p(k_{G^D})]^{\gamma}
\end{equation}
Specifying to the case of $z$-regular networks for simplicity, the previous equation takes a very simple form: 
\begin{equation}
p_z(\theta;\gamma, G^D) =  \left(1 - \frac{z^2}{2M_D} \right)^{\binom{\theta}{2}}
\left(\frac{\theta z^{2}}{2M_D}\right)^{\gamma - \theta}  \quad
p_z(\gamma;\gamma, G^D) =  \left(1 - \frac{z^2}{2M_D} \right)^{\binom{\theta}{2}}
\end{equation}
This form of the equation is interesting because it makes it easy to see the size dependence of the expected performance on both the average degree and size of the original $G_{TS}$: indeed in this case $2M=zM$, where $M$ here is the number of tasks in $G_D$ which can grow both by enlarging the size of $G_{TS}$ or by increasing its average degree or density, which both result in an increase of $M$.   
For a generic degree distribution with finite first and second moment, if we consider the three main factors in the probability as independent variables (which is a reasonable assumption for $\gamma\ll M$ or equivalently $M_{in}\ll M_D$), we can approximate the expressions above with the following:
\begin{equation}
p_k(\theta;\gamma, G^D) \simeq  \left(1 - \frac{\langle k^2 \rangle}{2M_D} \right)^{\binom{\theta}{2}}
\left(\frac{\theta \langle k \rangle ^{2}}{2M_D}\right)^{\gamma - \theta}  \quad
p_k(\gamma;\gamma, G^D) \simeq  \left(1 - \frac{\langle k^2 \rangle}{2M_D} \right)^{\binom{\gamma}{2}}
\end{equation} 
In Figure \ref{fig:ER-ig-densities.pdf} we show the results for $p_\gamma$, $\phi_\gamma$ and $\tilde{\phi}_\gamma$ for a set of Erdos-Renyi graphs with sizes $M$ between 10 and 150 nodes and densities $\rho$ between 0.2 and 0.6. 
In Figure \ref{fig:ER-fixed-degree.pdf} instead we show results obtained for Erdos-Renyi graphs of variable sizes where we kept the average degree fixed.
While in the latter case, we find that the size independence is broken (which is not surprising as increasing the network size at fixed average degree corresponds effectively to a network sparsification), we also find that qualitatively the results are the same as for the fixed density case. 
The expressions and the results shown above refer to networks with large $M$ because our aim is to investigate the scaling of effective capacity with increasing $M$ beyond the scale at which it is possible to simulate the neural networks. 
However, although our derivation is valid in the large $M$ limit, it is interesting to check whether it captures also{} the behaviour at very small $M$. 
In the simulations shown below, we start from the task structure graphs $G_{TS}$ and we build $G_D$ from them. Thus we do not control exactly the average degree nor the density of the resulting $G_D$ graphs, but we can fix the density of the input task structure graphs. 
In Figure \ref{fig:ER-fixed-density-ts.pdf} we show the resulting effective capacities obtained by doing this and computing numerically the resulting dependency graphs and using the expressions above to compute the effective capacities. We see that even in the small $M$ limit the analytical expressions capture well the functional forms obtained from simulations, up to an inversion of the trend when increasing from $M=2$ to $M=9$, which is likely driven by the increasing of the number of possible sets of size $\gamma$ when $M$ grows.


\clearpage
\section{Estimation of $G_D$'s degree sequence and distribution from $G_{TS}$}
In the main text we showed that under the minimal basis set scheme $\bar{G}_D$ coincides with the  $G_D$. This means that one can estimate the degree distribution of $G_D$ from $G_{TS}$ itself. Below we provide an estimation of this. 

\begin{figure}
\centering
\includegraphics[width=0.7\textwidth]{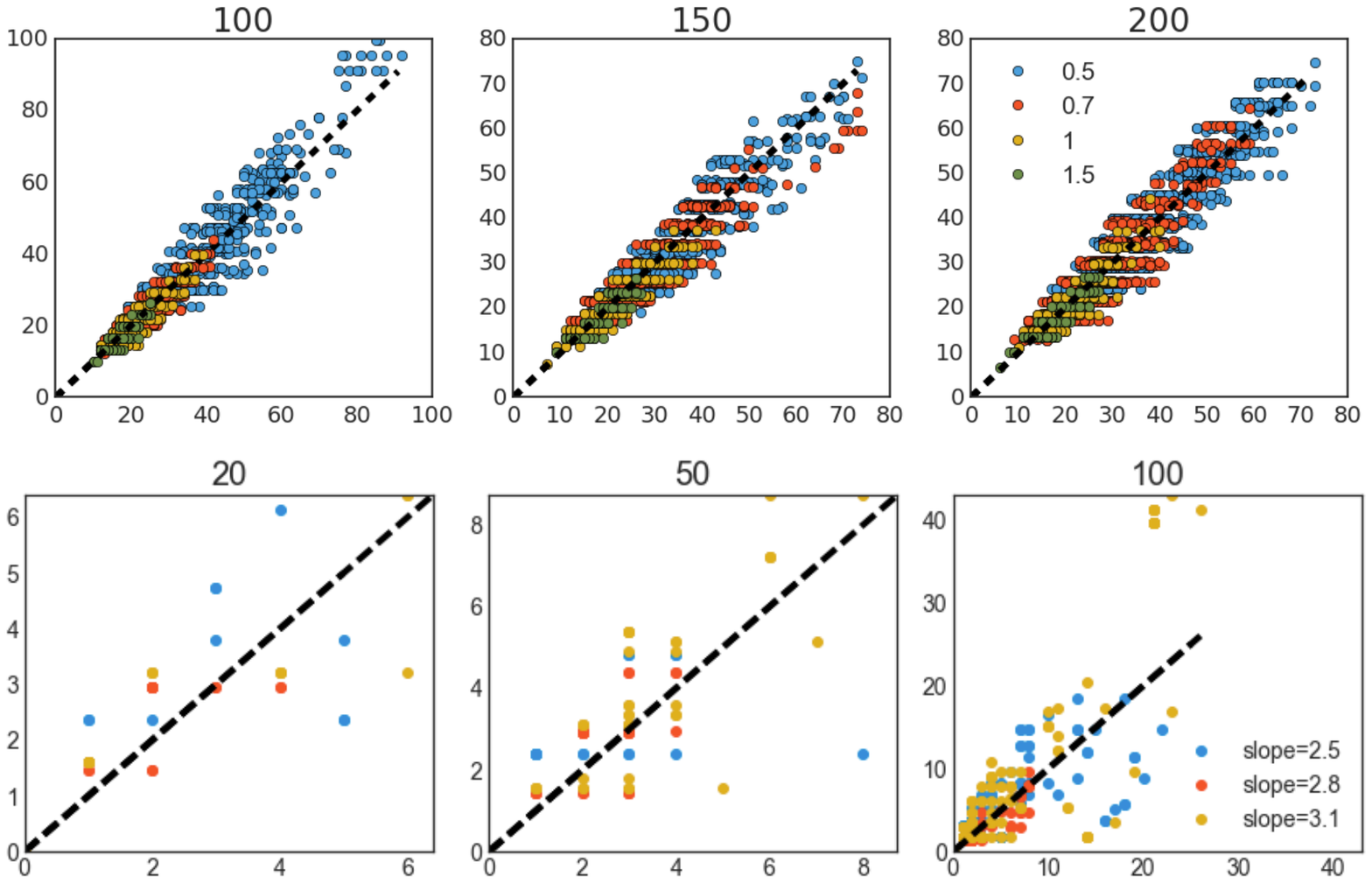}
\caption{\textbf{Estimated degrees in $G_D$.} We show here the comparison of the estimated degrees for individual nodes in the dependency graph with the actual corresponding degree in $G_D$. 
We show the results for two network topologies for the underlying task structure graph $G_{TS}$, exponential graphs with slopes ($0.5, 0.7, 1, 1.5$, top row) and scale-free graphs (with power law exponents $2.5,2.8,3.1$, bottom row), for various layer sizes of the underlying task structure graph $N=100$ (left), $N=150$ (center) and $N=200$ (right).}\label{fig:degrees}
\end{figure}

Consider an input-output pairing bipartite graph $G_{TS}$ with the same number of input and output nodes $N$, and denote its input degree distribution as $p(s)$ and the output degree distribution as $p(t)$. 
The line graph $\mathcal{L}(G_{TS})$ of $G_{TS}$ has node set $V_{\mathcal{L}(G_{TS})}$ equal to the edge set $E(G_{TS})$ of $G_{TS}$. 
Hence to each edge $e$ in $G_{TS}$, or equivalently node in the line graph $\mathcal{L}(G_{TS})$, we can associate its extremal nodes' degrees $(s_e, t_e)$.
The degree of $e$ is then $k^{\mathcal{L}(G_{TS})}_e = s+t-2$. 
Denoting the probability for an input node with degree $s$ to be linked to an output node with degree $t$ by $p_{st}$, we can use it to generate the degree distribution for the line graph $\mathcal{L}(G_{TS})$ as $p^{\mathcal{L}(G_{TS})}_k = \sum_{s,t} p_{st} \delta_{k,s+t-2}$.  
We then write the generating function for $p_{st}$ as $g_p(x,y) = \sum_{s,t=0}^\infty p_{st} x^s y^t$, which by substitution in the expression for $p^{\mathcal{L}(G_{TS})}_k$, yields $f(z)=\sum_{k=0}^\infty p^{\mathcal{L}(G_{TS})}_k z^k = \frac{1}{z^2} g_p(z,z)$ \cite{newman2009random}. 
The corresponding excess degree distributions become then
\begin{equation}
q_{st} = \frac{(t+1)p_{s,t+1}}{\langle t \rangle} \quad  
r_{st} = \frac{(s+1)p_{s+1,t}}{\langle s \rangle} 
\end{equation}which can be then rewritten in the standard generating function formalism. \\
In order to obtain an estimate $\tilde{k}^D_e$ for the actual degree $k^{D}_e$ in the dependency graph $G_D$ of a node $e$ in $\mathcal{L}(G_{TS})$ characterized by $(s,t)$, we need to calculate what is the contribution to the degree coming from the closure of open wedges, that is, how many second neighbours the node has. 
This calculation can be performed in a similar way to standard calculation of the number of second neighbours\cite{Newman:2010:NI:1809753}. 
In this case, however, we need to take care of the potential effect of the joint degree distribution $p_{st}$. 
The degree of node $e$ in the dependency graph $G_D$ is the sum of the degree of $e$ in $\mathcal{L}(G_{TS})$ and the number of second neighbours reached from the input excess edges and from the output excess edges: 
\begin{equation}
k^{D}_e  \simeq  (s+t-2) + (s-1)\sum_{t'} t' q_{s'=s,t'}  + (t-1)\sum_{s'} s' q_{s',t'=t}
  = (s-1) \frac{\langle t^2 \rangle}{\langle t\rangle} + (t-1) \frac{\langle s^2 \rangle}{\langle s\rangle}
\end{equation}
This result holds exactly for sparse graphs. However, we show that it also gives good results for graphs with intermediate densities.
In that case, however, we need to keep track of the possibility that input and output edges of a node in $\mathcal{L}(G_{TS})$ might connect to the same node. 
For a node $e \in V_{\mathcal{L}(G_{TS})}$ characterized by $(s,t)$, the expression for the degree correction takes the form 
$$\simeq (M-1)(s-1)(t-1) \sum_{s'=s,t'} \frac{t'}{M-1} \sum_{s',t'=t} r_{s',t} \frac{s'}{M-1} = \frac{(s-1)(t-1)(\langle s \rangle -1) (\langle t \rangle -1)}{M-1} $$.
Collecting all terms together, we arrive at:
\begin{equation}\label{eq:dependency-degree}
\bar{k}^{D}_e  \simeq  (s-1) \frac{\langle t^2 \rangle}{\langle t\rangle} + (t-1) \frac{\langle s^2 \rangle}{\langle s\rangle} - \frac{(s-1)(t-1)(\langle s \rangle -1) (\langle t \rangle -1)}{M-1}
 \end{equation}
Formally, we can write $\tilde{k}^{D} = \tilde{k}^{D}(s,t)$ and, similarly to above, this generates the degree distribution for the dependency graph $p^{G_{D}}_k = \sum_{s,t} p_{st} \delta_{k,\tilde{k}^D(s,t)}$. In practice, calculating the solution for this distribution can be cumbersome, especially when the $p_{st}$ does not have a well specified functional form. However, Eq. \ref{eq:dependency-degree} can directly be used to generate the degree sequence for $G_D$ from $G_{TS}$. 
Note also that the final expression for $\tilde{k}^{D}$ is written in terms of the first two moments of the task structure graph's degree distribution $p_{st}$, which plays a crucial role in the estimation of $\rho_\alpha$ of the $G_D$.
In Figure \ref{fig:degrees} we show the performance of Eq. \ref{eq:dependency-degree} for two different network topologies, exponential graphs ($p(k) \propto (1-e^{-\lambda}) e^{-\lambda*k})$) and scale-free graphs (). 
It is easy to see that in the exponential case, by virtue of the definiteness of the moments the estimation is quite accurate, while for scale-free graphs its accuracy is reduced due to the divergence of the second moment of the degree distribution (for slopes between 2 and 3). 

\end{document}